\documentclass[11pt,a4paper]{article}
\usepackage[utf8]{inputenc}
\usepackage[T1]{fontenc}
\usepackage{lmodern}
\usepackage[margin=0.72in]{geometry}
\usepackage{microtype}
\usepackage{graphicx,epsfig,epstopdf}
\usepackage{bm}
\usepackage{amssymb,amsmath,amsfonts,latexsym}
\usepackage{dcolumn}
\usepackage{multirow}
\usepackage{booktabs,longtable,array}
\usepackage{caption}
\usepackage{subcaption}
\usepackage{float}
\usepackage{tikz}
\usetikzlibrary{calc}
\usepackage[numbers,sort&compress]{natbib}
\usepackage{xcolor}
\usepackage{hyperref}
\hypersetup{
  colorlinks=true,
  linkcolor=blue,
  citecolor=blue,
  urlcolor=blue,
  filecolor=blue,
  allcolors=blue,
  pdfborder={0 0 0}
}
\makeatletter
\let\cat@comma@active\@empty
\let\merged@origcite\cite
\renewcommand{\cite}[1]{\ifmmode\text{\merged@origcite{#1}}\else\merged@origcite{#1}\fi}
\makeatother

\providecommand{\ccbar}{c\bar c}
\providecommand{\bbbar}{b\bar b}

\providecommand{\etac}{\eta_c}
\providecommand{\etab}{\eta_b}

\providecommand{\GeV}{\mathrm{GeV}}
\providecommand{\MeV}{\mathrm{MeV}}

\begin{document}

\title{A Unified Study of Hidden-Charm and Hidden-Bottom Mesons}

\author{Vikas Patel$^{1,2}$, Chetan Lodha$^{2}$, and A. K. Rai$^{2}$\\[0.4em]
\small $^{1}$Department of Physics, Uka Tarsadia University, Bardoli 394250, Gujarat, India\\
\small $^{2}$Department of Physics, Sardar Vallabhbhai National Institute of Technology, Surat 395007, Gujarat, India\\
\small Correspondence: \texttt{iamchetanlodha@gmail.com}}

\date{}
\maketitle

\begin{abstract}
We present a unified phenomenological analysis of hidden-charm and hidden-bottom mesons, treating the $c\bar c$ and $b\bar b$ sectors within a common screened-potential and QCD sum-rule framework.  The mass spectra are obtained from a screened Coulomb-plus-confining interaction with spin-dependent corrections, while the short-distance decay constants, annihilation widths and electromagnetic transition form factors are organized through two-point and three-point QCD sum rules.  The same calculated states are then used to study E1 and M1 radiative transitions, Regge trajectories and finite-spectrum thermodynamic observables constructed from excitation energies relative to the corresponding ground states.  The results are discussed together rather than as disconnected outputs: the mass spectra determine the state assignments, the Regge plots test the global level ordering, the radiative widths probe orbital and spin-flip overlap integrals, and the QCD sum-rule quantities connect the spectrum to current-coupled observables.  The comparison of the two hidden-flavour sectors shows the expected compression and stronger spin suppression in bottomonium, while charmonium remains more sensitive to fine-structure, radial-node and medium-related effects.  The resulting picture provides a coherent vacuum benchmark for $c\bar c$ and $b\bar b$ spectroscopy, radiative transitions and QCD sum-rule observables.
\end{abstract}

\section{Introduction}
Hidden-heavy quarkonia provide one of the most direct phenomenological laboratories for studying the transition between perturbative and nonperturbative QCD.  In charmonium ($c\bar c$) and bottomonium ($b\bar b$), the heavy constituent masses suppress many light-quark complications, while the bound states remain sensitive to confinement, spin-dependent forces, relativistic corrections and threshold effects.  The same spectrum therefore constrains several different pieces of the dynamics: the central potential fixes the gross level ordering, hyperfine and fine splittings test the spin-spin, spin-orbit and tensor interactions, decay constants probe the wave function at the origin, and radiative transitions test radial and angular overlap integrals.  For this reason, a useful hidden-flavour study should not be limited to mass spectra alone; it should connect spectroscopy with E1 and M1 transitions, Regge systematics and QCD sum-rule observables within a common framework \cite{pdg,M. B. Voloshin,N.Brambilla}.

The experimental history of the $c\bar c$ and $b\bar b$ sectors also motivates such a connected analysis.  The discoveries of the $J/\psi$ and $\Upsilon$ families established narrow hidden-heavy resonances as precision probes of the strong interaction \cite{ExpJpsiBNL,ExpJpsiSLAC,ExpUpsilonE288}.  Subsequent measurements of spin-singlet, orbital and radially excited states, including $\eta_c$, $\eta_c(2S)$, $\chi_{c1}(3872)$, $\psi_2(3823)$, $\psi_3(3842)$, $\eta_b(1S)$ and $h_b(1P,2P)$, have made the spectroscopy much richer and have sharpened the need for reliable assignments of both conventional and threshold-sensitive levels \cite{ExpEtaC,ExpEtaC2S,ExpX3872Belle,ExpPsi3823BESIII,ExpX3842LHCb,ExpEtaBBaBar,ExpHBelle}.  Representative higher states used as orientation points in the present discussion are collected in Table~\ref{Table:intro table}.  The table is not intended to replace the detailed comparison in Sec.~\ref{sec:results}; rather, it identifies the experimental landscape against which the calculated hidden-charm and hidden-bottom spectra are judged.

\begin{table*}[!htb]
\caption{Selected $c\bar c$- and $b\bar b$-sector experimental states used for orientation. Masses are aligned with the PDG 2025 listings and quoted in MeV \cite{pdg}.}
\label{Table:intro table}
\centering
\renewcommand{\arraystretch}{1.08}
\resizebox{\textwidth}{!}{%
\begin{tabular}{lllll}
\toprule
PDG state & Common name & $J^{PC}$ & PDG mass (MeV) & Main observed production or decay mode \\
\midrule
$\psi(3770)$ & -- & $1^{--}$ & $3773.7\pm0.7$ & $e^+e^-\to D\bar D$ \\
$\psi_2(3823)$ & -- & $2^{--}$ & $3823.51\pm0.34$ & $B\to K\gamma\chi_{c1}$, $\pi^+\pi^-J/\psi$ \\
$\psi_3(3842)$ & -- & $3^{--}$ & $3842.71\pm0.16\pm0.12$ & $D\bar D$ \\
$\chi_{c1}(3872)$ & $X(3872)$ & $1^{++}$ & $3871.64\pm0.06$ & $B\to K\pi^+\pi^-J/\psi$ \\
$\chi_{c0}(3915)$ & $X(3915)$ & $0^{++}$ & $3922.1\pm1.8$ & $\gamma\gamma\to\omega J/\psi$ \\
$\chi_{c2}(3930)$ & -- & $2^{++}$ & $3922.5\pm1.0$ & $\gamma\gamma\to D\bar D$ \\
$\psi(4040)$ & -- & $1^{--}$ & $4040\pm4$ & $e^+e^-\to$ hadrons \\
$\chi_{c1}(4140)$ & $X(4140)$ & $1^{++}$ & $4146.5\pm3.0$ & $B^+\to J/\psi\phi K^+$ \\
$\psi(4160)$ & -- & $1^{--}$ & $4191\pm5$ & $e^+e^-\to$ hadrons \\
$\psi(4230)$ & $Y(4230)$ & $1^{--}$ & $4222.2\pm2.4$ & $e^+e^-\to\pi^+\pi^-J/\psi$, $\omega\chi_{c0}$ \\
$\chi_{c1}(4274)$ & $X(4274)$ & $1^{++}$ & $4286^{+8}_{-9}$ & $B^+\to J/\psi\phi K^+$ \\
$\psi(4360)$ & $Y(4360)$ & $1^{--}$ & $4374\pm7$ & $e^+e^-\to\pi^+\pi^-\psi(2S)$ \\
$\psi(4415)$ & -- & $1^{--}$ & $4415\pm5$ & $e^+e^-\to$ hadrons \\
$\psi(4660)$ & $Y(4660)$ & $1^{--}$ & $4623\pm10$ & $e^+e^-\to\pi^+\pi^-\psi(2S)$ \\
\midrule
$\Upsilon(4S)$ & -- & $1^{--}$ & $10579.4\pm1.2$ & $e^+e^-\to B\bar B$ \\
$\Upsilon(10753)$ & -- & $1^{--}$ & $10752.7\pm5.9{}^{+0.7}_{-1.1}$ & $e^+e^-\to\Upsilon(nS)\pi^+\pi^-$ \\
$\Upsilon(10860)$ & -- & $1^{--}$ & $10885.2^{+2.6}_{-1.6}$ & $e^+e^-\to$ hadrons \\
$\Upsilon(11020)$ & -- & $1^{--}$ & $11000\pm4$ & $e^+e^-\to$ hadrons \\
\bottomrule
\end{tabular}}
\renewcommand{\arraystretch}{1.0}
\end{table*}

On the theoretical side, potential models remain valuable because they encode the dominant features of heavy-quark dynamics in a transparent way.  Cornell-type interactions and their screened or relativized extensions combine a short-distance Coulomb term with a long-distance confining term, while spin-dependent corrections resolve the physical $J^{PC}$ multiplets \cite{Eichten1978Cornell,Eichten1980Comparison,E.Eichten2,S. Godfrey2}.  Bottomonium, being more compact, provides a stringent check on the short-distance part of the interaction and on the heavy-quark expansion.  Charmonium, in contrast, is more sensitive to relativistic corrections, open-charm thresholds and screening effects.  Treating both sectors in parallel is therefore useful: agreement in $b\bar b$ tests the stability of the short-distance normalization, whereas the $c\bar c$ spectrum tests how well the same framework accommodates stronger relativistic and threshold sensitivity.

A complete phenomenological description also requires observables beyond masses.  E1 transitions mainly connect states with different orbital angular momentum and the same total spin, so their widths are controlled by photon-energy factors, angular coefficients and radial matrix elements.  M1 transitions mainly connect spin-singlet and spin-triplet states with the same orbital angular momentum and are more sensitive to spin-flip overlaps and wave-function orthogonality.  Decay constants and annihilation widths probe short-distance current couplings, while transition form factors organize the momentum dependence of electromagnetic and weak-current matrix elements.  QCD sum rules provide an independent language for these short-distance quantities by relating pole residues and form factors to operator-product expansions, spectral densities, Borel transformations and continuum thresholds \cite{Shifman1979Foundations,Reinders1985QCDSR,Y.B.Dai,Y.R. Liu,Veliev2011Vector}.  Lattice-QCD, nonrelativistic QCD (NRQCD), potential NRQCD (pNRQCD) and light-cone sum-rule studies provide complementary benchmarks for the same quantities \cite{Donald2012JpsiLattice,Delaney2024LatticeRadiative,Guo2020LCSRJpsiEtaC,Bodwin1995NRQCD,Brambilla2006M1}.

The present article follows this multi-observable strategy.  The $c\bar c$ and $b\bar b$ spectra are obtained from a screened Coulomb-plus-confining potential supplemented by spin-dependent interactions.  The resulting states are then used to study spin-averaged and spin-resolved masses, Regge trajectories, E1 and M1 radiative widths, decay constants, annihilation widths, transition form factors and finite-spectrum thermodynamic indicators.  A closely related unified heavy-heavy analysis has recently been carried out for the unequal-mass $B_c$ system, where spectroscopy, transition form factors, weak decays and radiative widths were treated in a common framework \cite{Patel:2026prg}.  That work is useful as a methodological analogue because it demonstrates how mass spectra and current-induced observables can be discussed together; however, the present study is not a $B_c$ calculation and keeps the focus on hidden-flavour $c\bar c$ and $b\bar b$ states.  The thermodynamic quantities are constructed only from the discrete zero-temperature quarkonium spectrum; they are included as diagnostics of level spacing and state multiplicity, not as a full finite-temperature QCD medium calculation.  This distinction is important because recent spectrum-based thermodynamic studies and hot-medium susceptibility analyses address related but physically different questions \cite{AbuShadyFathAllah2025,SamantaBroniowski2026}.

Regge phenomenology is used here as a global test of the calculated spectrum.  If the screened potential gives a coherent description of the radial and orbital excitations, the states should approximately organize into linear patterns in the $(M^2,J)$ and $(M^2,n_r)$ planes.  Such trajectories do not replace the detailed mass calculation, but they summarize how the same Hamiltonian arranges the low-lying and excited states.  Related Regge and heavy-hadron studies show that slopes, intercepts and deviations from linearity can be useful diagnostics of excitation patterns across mesons, baryons and multiquark systems \cite{Purohit:2022mwu,Kher:2022gbz,Chaturvedi:2022pmn,lodha:2025snn,Lodha:2026iww,Lodha:2024qby,Patel:2025rsf,Jakhad:2025zos}.  Recent applications to beauty bound states and compact charm-strange tetraquarks also show how spectral assignments can be connected with decay thresholds, rearrangement channels and Regge trajectories in neighbouring heavy-hadron sectors \cite{Lodha:2026iww,Lodha:2026uwi}.  Broader multiquark applications, such as hypercentral studies of singly charmed pentaquarks, further motivate using trajectory and mass-systematics arguments with clear separation between conventional quarkonium and multiquark interpretations \cite{Rathod:2026bph}.

Table~\ref{tab:theory_studies_intro} gives an introductory map of representative theoretical studies relevant to the present analysis.  The table is included to clarify the roles of the main comparison frameworks: potential and screened-potential calculations address mass ordering and splittings; lattice-QCD studies supply nonperturbative benchmarks; QCD sum rules determine pole residues, decay constants and form factors; effective-field-theory approaches organize annihilation and radiative corrections; and spectrum-based thermodynamic analyses clarify how thermodynamic indicators can be built from discrete levels.  These references are used throughout Sec.~\ref{sec:results} to place the present hidden-flavour results in context.

{\footnotesize
\setlength{\LTleft}{0pt}
\setlength{\LTright}{0pt}
\setlength{\tabcolsep}{3pt}
\renewcommand{\arraystretch}{1.18}
\begin{longtable}{|>{\raggedright\arraybackslash}p{0.235\textwidth}|>{\centering\arraybackslash}p{0.105\textwidth}|>{\raggedright\arraybackslash}p{0.285\textwidth}|>{\raggedright\arraybackslash}p{0.285\textwidth}|}
\caption{Introductory map of representative theoretical and adjacent heavy-hadron studies relevant to the hidden-charm and hidden-bottom analysis.}\label{tab:theory_studies_intro}\\
\hline
\textbf{Study} & \textbf{Sector} & \textbf{Theoretical framework} & \textbf{Main quantities addressed} \\
\hline
\endfirsthead
\caption[]{Introductory map of representative theoretical and adjacent heavy-hadron studies relevant to the hidden-charm and hidden-bottom analysis.}\\
\hline
\textbf{Study} & \textbf{Sector} & \textbf{Theoretical framework} & \textbf{Main quantities addressed} \\
\hline
\endhead
\hline
\multicolumn{4}{r}{\emph{Continued on next page}}\\
\hline
\endfoot
\hline
\endlastfoot
Cornell quarkonium calculations \cite{Eichten1978Cornell,Eichten1980Comparison} & $c\bar c$, $b\bar b$ & Coulomb-plus-linear confinement with spin-dependent corrections & Low-lying spectra, $1S$--$1P$--$2S$ ordering, spin splittings and the separation of Coulombic and confining dynamics. \\
\hline
Eichten--Godfrey--Mahlke--Rosner review \cite{E.Eichten2} & $c\bar c$, $b\bar b$ & Quarkonium review with threshold and coupled-channel emphasis & Conventional assignments, open-flavour threshold effects, coupled-channel mass shifts and experimental status of heavy-quarkonium levels. \\
\hline
Godfrey--Isgur \cite{S. Godfrey2} & $c\bar c$, $b\bar b$ & Relativized constituent-quark Hamiltonian & Spin-resolved spectra, relativistic corrections, fine splittings and orbital-multiplet ordering. \\
\hline
Devlani--Kher--Rai \cite{N. Devlani} & $c\bar c$, $b\bar b$ & Nonrelativistic quark model with spin-dependent interaction & S-, P-, D- and F-wave masses together with hyperfine, spin-orbit and tensor splittings. \\
\hline
Kher--Chaturvedi--Devlani--Rai and Purohit--Jakhad--Rai \cite{Kher:2022gbz,Purohit:2022mwu} & $c\bar c$, $b\bar b$ & Cornell and modified-Yukawa quarkonium Hamiltonians & Recent quarkonium mass systematics, radial excitations, orbital excitation patterns and bottomonium benchmarks. \\
\hline
Koma--Koma--Wittig \cite{Y. Koma} & $c\bar c$, $b\bar b$ & Lattice determination of spin-dependent potentials & Nonperturbative spin-orbit, tensor and hyperfine force components extracted from lattice gauge theory. \\
\hline
Ebert--Faustov--Galkin \cite{D. Ebert} & $c\bar c$, $b\bar b$ & Relativistic quasipotential approach with QCD-motivated interaction & Full hidden-flavour spectra, relativistic mass shifts and Regge-trajectory behavior. \\
\hline
Buchmuller--Tye and potential-review studies \cite{Buchmuller1981Quarkonia,Lucha1991BoundStates} & $c\bar c$, $b\bar b$ & Phenomenological heavy-quark potentials and bound-state review framework & Potential parameters constrained by level spacings, hyperfine splittings, leptonic widths and global quarkonium systematics. \\
\hline
Li--Chao and Deng \textit{et al.} \cite{B.Q. Li1,W.J. Deng} & $c\bar c$ & Screened and nonrelativistic quark models with electromagnetic multipoles & Higher charmonium masses, screened-confinement effects and E1/M1 radiative-transition widths. \\
\hline
Barnes--Godfrey--Swanson \cite{T. Barnes} & $c\bar c$ & Quark-model survey of higher charmonia and decay systematics & Excited-charmonium assignments, strong-decay modes and radiative-transition patterns. \\
\hline
Kawanai--Sasaki \cite{T. Kawanai} & $c\bar c$ & Lattice-QCD determination of the heavy-quark potential & Charmonium interquark potential and central-force information from lattice simulations. \\
\hline
Godfrey--Moats, Segovia \textit{et al.}, and Lu--Anwar--Zou \cite{S. Godfrey1,J. Segovia,Y. Lu} & $b\bar b$ & Relativized, coupled-channel and constituent-quark descriptions & Bottomonium masses, E1/M1 widths, annihilation widths, high-lying levels and threshold-sensitive assignments. \\
\hline
Dudek \textit{et al.}, Liu \textit{et al.}, and HPQCD bottomonium studies \cite{Dudek2008CharmoniumLattice,Liu2012CharmoniumLattice,Gray2005UpsilonLattice,Dowdall2012UpsilonSpectrum} & $c\bar c$, $b\bar b$ & Lattice-QCD and lattice NRQCD spectroscopy & Excited multiplets, spin assignments, radial spacings, hyperfine splittings and heavy-quark-limit spectroscopy. \\
\hline
SVZ/Reinders QCD sum rules and thermal-vector extensions \cite{Shifman1979Foundations,Reinders1985QCDSR,Y.B.Dai,Y.R. Liu,Veliev2011Vector,Kim2023Momentum,Kim2023Rotating,Kim2025SpinThermal} & $c\bar c$, $b\bar b$ & Current correlators, OPE, Borel transformation, continuum subtraction and medium-modified sum rules & Pole residues, decay constants, condensate contributions, thermal residues, finite-momentum effects and spin decomposition. \\
\hline
HPQCD $J/\psi$ calculation and radiative form-factor studies \cite{Donald2012JpsiLattice,Delaney2024LatticeRadiative,Guo2020LCSRJpsiEtaC} & Mainly $c\bar c$ & Lattice-QCD and light-cone sum-rule transition form factors & $J/\psi$ leptonic width, $J/\psi\to\eta_c\gamma$ normalization, M1 form factors and radiative-transition uncertainties. \\
\hline
NRQCD and pNRQCD decay analyses \cite{Bodwin1995NRQCD,Brambilla2006M1,Petrelli1998NLO,Brambilla2003Inclusive,Brambilla2020StronglyCoupled} & $c\bar c$, $b\bar b$ & Factorization with short-distance coefficients and long-distance matrix elements & Inclusive annihilation widths, short-distance coefficients, M1 corrections and heavy-quark spin-symmetry constraints. \\
\hline
$B_c$ unified spectroscopy and decay analysis \cite{Patel:2026prg} & $B_c$ & Screened-potential spectroscopy combined with QCD sum-rule form factors and weak/radiative decay formalism & Unequal-mass heavy-heavy spectrum, transition form factors, weak decay modes and radiative widths; used here only as a methodological analogue, not as a hidden-flavour input. \\
\hline
Beauty and charm-strange multiquark studies \cite{Lodha:2026iww,Lodha:2026uwi} & $B$, $bq\bar q\bar q$, $cs\bar s\bar s$ & Screened-potential, decay-dynamics and Regge-trajectory analyses with threshold and rearrangement-channel inputs & Open-beauty spectra, decay dynamics, compact tetraquark mass patterns, rearrangement decays and Regge behaviour in neighbouring heavy-hadron sectors. \\
\hline
Hypercentral pentaquark study \cite{Rathod:2026bph} & Singly charmed pentaquark & Hypercentral constituent-quark model & Multiquark mass systematics in the charmed sector; included as broader context for distinguishing conventional hidden-flavour quarkonium from multiquark spectroscopy. \\
\hline
Generalized fractional Klein--Gordon study \cite{AbuShadyFathAllah2025} & $c\bar c$, $b\bar b$ & Relativistic wave-equation solution with screened Kratzer/Yukawa-type interactions & Screened relativistic mass spectra and finite-spectrum thermodynamic indicators for heavy mesons. \\
\hline
Hot-medium magnetic-susceptibility study \cite{SamantaBroniowski2026} & Hot hadronic medium & Hadron-resonance-gas and quark--meson framework constrained by lattice-QCD susceptibilities & Magnetic susceptibility, baryon and strangeness susceptibilities, and the distinction between bulk thermal-medium response and spectrum-derived thermodynamic indicators. \\
\hline
\end{longtable}
\renewcommand{\arraystretch}{1.0}
}

The aim of the present article is to provide a coherent hidden-charm and hidden-bottom analysis in which spectroscopy, radiative transitions and QCD sum-rule observables are treated as connected probes of the same calculated states.  The article is organized as follows.  Section~2 combines the theoretical framework and numerical implementation: it introduces the screened-potential Hamiltonian, spin-dependent interaction, short-distance decay constants, QCD sum-rule residues and transition form factors, annihilation widths, E1 and M1 transitions, finite-spectrum thermodynamic observables, Regge construction, input parameters, Borel windows and uncertainty treatment.  Section~3 contains the spin-averaged and spin-resolved spectra, Regge trajectories, radiative transitions, thermodynamic observables, decay constants, annihilation widths, transition form factors and comparisons with representative theoretical and experimental studies.  Finally, Sec.~4 merges the numerical synthesis with the main conclusions, so the final section does not repeat the Results and Discussion but summarizes the physical picture emerging from all observables.

\section{Theoretical framework and numerical implementation}\label{sec:framework}

This section gives the complete methodological setup used in the hidden-charm and hidden-bottom analysis.  The bound-state sector is based on a screened Coulomb-plus-confining interaction with relativistic kinetic corrections, Gaussian variational wave functions and spin-dependent forces.  The same calculated states then provide the mass and wave-function input for radiative transitions, Regge trajectories and finite-spectrum thermodynamic indicators.  The current-coupled decay observables are handled through QCD sum-rule residues and three-point form factors, so the decay constants, annihilation widths and transition form factors are tied to Borel stability, continuum thresholds and pole dominance.  For clarity, the formal definitions and the numerical implementation are presented together rather than in separate sections: the equations specify the framework, while the input table, Borel windows and uncertainty rules state how the framework is evaluated.

\subsection{Screened-potential Hamiltonian and variational basis}

For a hidden-heavy quarkonium state $Q\bar Q$, with $Q=c$ or $b$, the starting point is the semi-relativistic Hamiltonian
\begin{equation}
H=\sqrt{{\bf p}^{2}+m_Q^2}+\sqrt{{\bf p}^{2}+m_{\bar Q}^2}+V(r),
\label{Eq:hamiltonian}
\end{equation}
where ${\bf p}$ is the relative momentum and $r$ is the quark-antiquark separation.  In the present hidden-flavour applications $m_Q=m_{\bar Q}=m_c$ for charmonium and $m_Q=m_{\bar Q}=m_b$ for bottomonium.  After subtracting the constituent rest masses, the kinetic-energy operator is expanded as
\begin{eqnarray}
T_{\rm rel} &=& \frac{{\bf p}^{2}}{2}\left(\frac{1}{m_Q}+\frac{1}{m_{\bar Q}}\right)
-\frac{{\bf p}^{4}}{8}\left(\frac{1}{m_Q^{3}}+\frac{1}{m_{\bar Q}^{3}}\right)
+\frac{{\bf p}^{6}}{16}\left(\frac{1}{m_Q^{5}}+\frac{1}{m_{\bar Q}^{5}}\right) \nonumber\\
&&-\frac{5{\bf p}^{8}}{128}\left(\frac{1}{m_Q^{7}}+\frac{1}{m_{\bar Q}^{7}}\right)
+\frac{7{\bf p}^{10}}{256}\left(\frac{1}{m_Q^{9}}+\frac{1}{m_{\bar Q}^{9}}\right).
\label{Eq:cornel}
\end{eqnarray}
The expansion is used as a controlled phenomenological approximation for heavy-heavy systems; it is more rapidly convergent in bottomonium and gives the leading relativistic corrections needed in charmonium.

The central interaction is written as
\begin{equation}
V(r)=V^{(0)}(r)+\left(\frac{1}{m_Q}+\frac{1}{m_{\bar Q}}\right)V^{(1)}(r)+{\cal O}\left(\frac{1}{m^2}\right),
\label{Eq:potential}
\end{equation}
with
\begin{equation}
V^{(0)}(r)=V_v(r)+V_s(r),
\label{Eq:eqation01}
\end{equation}
where the vector and scalar pieces are chosen as
\begin{equation}
V_v(r)=-\frac{4}{3}\frac{\alpha_s}{r},
\qquad
V_s(r)=\frac{A}{\xi}\left(1-e^{-\xi r}\right)+V_0.
\end{equation}
The vector term represents one-gluon exchange, while the scalar term gives confinement that is approximately linear at moderate distances and screened at large distances.  Screening is important for high radial and orbital excitations, where the states are more extended and are expected to feel the onset of open-flavour thresholds and coupled-channel effects \cite{E.Eichten2,A.K. Rai,N. Devlani,S. Godfrey,B.Q. Li1,S. Godfrey2,W.J. Deng,E.J. Eichte3,E.van Beveren,Vikas Patel}.  The leading short-distance correction is retained in the form
\begin{equation}
\begin{aligned}
V(r)=&-\frac{4}{3}\frac{\alpha_s}{r}+\frac{A}{\xi}\left(1-e^{-\xi r}\right)+V_0
+\left(\frac{1}{m_Q}+\frac{1}{m_{\bar Q}}\right)V^{(1)}(r)
+{\cal O}\left(\frac{1}{m^2}\right),
\end{aligned}
\label{eqn:6}
\end{equation}
where
\begin{equation}
V^{(1)}(r)=-\frac{C_FC_A\alpha_s^2}{4r^2},
\qquad C_F=\frac{4}{3},\quad C_A=3.
\label{eqn:7}
\end{equation}
This term improves the short-distance part of the interaction beyond the static central potential \cite{Y. Koma,N.Brambilla}.

The radial eigenvalues are obtained variationally using normalized Gaussian basis functions.  In coordinate space we use
\begin{equation}
\begin{aligned}
R_{nl}(\beta,r)=&\,\beta^{3/2}
\left[\frac{2(n-1)!}{\Gamma(n+l+1/2)}\right]^{1/2}
(\beta r)^l e^{-\beta^2r^2/2}
L_{n-1}^{l+1/2}(\beta^2r^2),
\end{aligned}
\label{eqn:8}
\end{equation}
and in momentum space
\begin{equation}
\begin{aligned}
R_{nl}(\beta,p)=&\,\frac{(-1)^n}{\beta^{3/2}}
\left[\frac{2(n-1)!}{\Gamma(n+l+1/2)}\right]^{1/2}
\left(\frac{p}{\beta}\right)^l e^{-p^2/(2\beta^2)}
L_{n-1}^{l+1/2}\left(\frac{p^2}{\beta^2}\right).
\end{aligned}
\label{eqn:9}
\end{equation}
Here $\beta$ is the state-dependent variational width, $L_{n-1}^{l+1/2}$ denotes the associated Laguerre polynomial and $\Gamma$ is Euler's gamma function.  Because the Hamiltonian in Eq.~\eqref{Eq:hamiltonian} is treated with the relativistic kinetic-energy expansion of Eq.~\eqref{Eq:cornel}, the width is fixed by the stationarity of the full expectation value rather than by the lowest-order nonrelativistic virial relation alone:
\begin{equation}
\frac{\partial}{\partial\beta}\langle H\rangle_{nl\beta}=0,
\qquad
\left\langle {\bf p}\cdot\frac{\partial T_{\rm rel}({\bf p})}{\partial {\bf p}}\right\rangle
=
\left\langle r\frac{dV}{dr}\right\rangle .
\label{eqn:virial-theorem}
\end{equation}
The second equality is the corresponding generalized virial condition for a Hamiltonian of the form $T({\bf p})+V(r)$.  If only the leading ${\bf p}^2/(2\mu)$ term is retained it reduces to the familiar nonrelativistic form, but in the numerical minimization the expectation value of the expanded kinetic operator in Eq.~\eqref{Eq:cornel} is used.  Thus each radial and orbital state has its own optimized wave function.  The spin-averaged $S$-wave mass is written as
\begin{equation}
M_{SA}=M_P+\frac{3}{4}(M_V-M_P),
\label{eqn:10}
\end{equation}
where $M_P$ and $M_V$ denote the pseudoscalar and vector masses in the same radial multiplet.  For higher orbital multiplets the center-weighted mass is
\begin{equation}
M_{CW,n}=\frac{\sum_J(2J+1)M_{nJ}}{\sum_J(2J+1)}.
\label{eqn:11}
\end{equation}
These spin-averaged quantities are the bridge between the central Hamiltonian and the resolved fine-structure spectrum.

\subsection{Spin-dependent interaction and wave function at the origin}

The central spectrum is resolved into physical $J^{PC}$ states by adding the spin-dependent interaction
\begin{eqnarray}
V_{\rm SD}(r)&=&
\left(\frac{{\bf L}\cdot{\bf S}_{Q}}{2m_Q^2}+\frac{{\bf L}\cdot{\bf S}_{\bar Q}}{2m_{\bar Q}^2}\right)
\left[-\frac{1}{r}\frac{dV^{(0)}(r)}{dr}+\frac{8\alpha_s}{3r^3}\right]
+\frac{4\alpha_s}{3m_Qm_{\bar Q}}\frac{{\bf L}\cdot{\bf S}}{r^3} \nonumber\\
&&+\frac{32\pi\alpha_s}{9m_Qm_{\bar Q}}({\bf S}_{Q}\cdot{\bf S}_{\bar Q})\delta^{(3)}({\bf r})
+\frac{4\alpha_s}{3m_Qm_{\bar Q}}
\frac{3({\bf S}_{Q}\cdot{\bf n})({\bf S}_{\bar Q}\cdot{\bf n})-({\bf S}_{Q}\cdot{\bf S}_{\bar Q})}{r^3},
\label{eq:spinhyperfine}\label{Eq:spinhyperfine}
\end{eqnarray}
where ${\bf S}={\bf S}_Q+{\bf S}_{\bar Q}$ and ${\bf n}={\bf r}/r$.  The four terms describe, respectively, the spin-orbit contribution associated with the scalar-vector potential mixture, the symmetric spin-orbit term from one-gluon exchange, the contact hyperfine interaction and the tensor interaction.  Their expectation values generate the hyperfine splittings of the $S$ waves and the fine splittings of the $P$-, $D$- and $F$-wave multiplets.

Short-distance observables depend strongly on the wave function at the origin.  We therefore use the spin-corrected radial wave function
\begin{equation}
R_{nJ}(0)=R(0)\left[1+(SF)_J\frac{\langle\varepsilon_{SD}\rangle_{nJ}}{M_{SA}}\right],
\label{eqn:radial-wavefunction-correction}
\end{equation}
where $(SF)_J$ is the spin factor appropriate to the state and $\langle\varepsilon_{SD}\rangle_{nJ}$ is the spin-dependent energy correction.  The spin-averaged value entering the correction is
\begin{equation}
R(0)=\frac{R_P+3R_V}{4}.
\label{eqn:radial-origin-average}
\end{equation}
This prescription keeps the decay constants and annihilation widths tied to the same fine-structure calculation used for the mass tables.

\subsection{QCD sum-rule residues and decay constants}

The QCD sum-rule part of the framework is used for the current-coupled decay observables.  For a hidden-heavy state $H=Q\bar Q$, the pseudoscalar and vector interpolating currents are
\begin{equation}
j_5(x)=\bar Q(x)i\gamma_5Q(x),\qquad
j_\mu(x)=\bar Q(x)\gamma_\mu Q(x).
\end{equation}
The corresponding two-point correlation functions are
\begin{equation}
\Pi_5(q^2)=i\int d^4x\,e^{iq\cdot x}\langle0|T\{j_5(x)j_5^\dagger(0)\}|0\rangle,
\label{eq:pi5}
\end{equation}
\begin{equation}
\Pi_{\mu\nu}(q)=i\int d^4x\,e^{iq\cdot x}\langle0|T\{j_\mu(x)j_\nu^\dagger(0)\}|0\rangle
=\left(-g_{\mu\nu}+\frac{q_\mu q_\nu}{q^2}\right)\Pi_V(q^2)+\frac{q_\mu q_\nu}{q^2}\Pi_L(q^2).
\label{eq:piv}
\end{equation}
The decay constants are defined by
\begin{equation}
\langle0|j_5|P(q)\rangle=\frac{f_PM_P^2}{2m_Q},
\qquad
\langle0|j_\mu|V(q,\lambda)\rangle=f_VM_V\varepsilon_\mu^{(\lambda)}.
\label{eq:decay_constants_def}
\end{equation}
After the Borel transformation and continuum subtraction, the working two-point sum rules can be written compactly as
\begin{equation}
\frac{f_P^2M_P^4}{4m_Q^2}e^{-M_P^2/M^2}
=\int_{4m_Q^2}^{s_0^P}ds\,\rho_5^{\rm pert}(s)e^{-s/M^2}+\widehat{\Pi}_5^{\rm cond}(M^2),
\label{eq:sr_p}
\end{equation}
\begin{equation}
f_V^2M_V^2e^{-M_V^2/M^2}
=\int_{4m_Q^2}^{s_0^V}ds\,\rho_V^{\rm pert}(s)e^{-s/M^2}+\widehat{\Pi}_V^{\rm cond}(M^2).
\label{eq:sr_v}
\end{equation}
The thresholds $s_0^P$ and $s_0^V$ separate the ground-state pole from the higher-state continuum, while $\widehat{\Pi}^{\rm cond}$ contains the nonperturbative condensate contribution retained in the operator product expansion.  The parameter $M^2$ is the Borel mass squared introduced by the Borel transformation; it is varied inside the Borel window quoted in Table~\ref{tab:inputs}.  The pole mass associated with a given channel is obtained from the logarithmic derivative
\begin{equation}
M_H^2=\frac{\frac{d}{d(-1/M^2)}{\cal L}_H(M^2,s_0)}{{\cal L}_H(M^2,s_0)},
\qquad
{\cal L}_H(M^2,s_0)=\int_{4m_Q^2}^{s_0}ds\,\rho_H^{\rm pert}(s)e^{-s/M^2}+\widehat{\Pi}_H^{\rm cond}(M^2).
\label{eq:mass_derivative}
\end{equation}
The accepted Borel windows are chosen by requiring simultaneous pole dominance, OPE convergence and residual stability of the extracted decay constant.  These requirements are imposed separately in the charmonium and bottomonium sectors because the bottom-quark mass shifts the stable Borel domain upward and suppresses condensate corrections more strongly.

\subsection{Annihilation widths from QCD sum-rule residues}

The residues obtained from the two-point sum rules enter the short-distance annihilation widths.  For pseudoscalar quarkonia we use
\begin{equation}
\Gamma(P\to\gamma\gamma)=\frac{4\pi\alpha_{\rm em}^2 e_Q^4 f_P^2}{M_P}
\left[1+\frac{\pi^2-20}{3}\frac{\alpha_s}{\pi}\right],
\label{eq:P_gammagamma}
\end{equation}
\begin{equation}
\Gamma(P\to gg)=\frac{8\pi}{3}\frac{\alpha_s^2f_P^2}{M_P}.
\label{eq:P_gg}
\end{equation}
For vector quarkonia, the dileptonic width is
\begin{equation}
\Gamma(V\to\ell^+\ell^-)=\frac{4\pi\alpha_{\rm em}^2e_Q^2f_V^2}{3M_V}
\left(1+\frac{2m_\ell^2}{M_V^2}\right)
\left(1-\frac{4m_\ell^2}{M_V^2}\right)^{1/2},
\label{eq:V_ll}
\end{equation}
and the three-gluon width is estimated through
\begin{equation}
\Gamma(V\to ggg)=\Gamma(V\to e^+e^-)
\frac{10(\pi^2-9)}{81\pi e_Q^2}\frac{\alpha_s^3}{\alpha_{\rm em}^2}.
\label{eq:V_ggg}
\end{equation}
These expressions make explicit why decay constants and strong-coupling uncertainties dominate the annihilation-width error budget.

\subsection{Electromagnetic transition form factors from three-point sum rules}

The electromagnetic transition form factors are extracted from three-point current correlators.  For a pseudoscalar quarkonium state coupled to one on-shell and one off-shell photon, the correlator is
\begin{equation}
\Pi_{\mu\nu}(p,q)=i^2\int d^4x\,d^4y\,e^{iq\cdot x+ip\cdot y}
\langle0|T\{j_\mu^{\rm em}(x)j_\nu^{\rm em}(y)j_5^\dagger(0)\}|0\rangle,
\label{eq:three_p_gamma}
\end{equation}
with
\begin{equation}
\Pi_{\mu\nu}(p,q)=i\epsilon_{\mu\nu\alpha\beta}p^\alpha q^\beta\Pi_P(p^2,q^2,(p+q)^2).
\end{equation}
The form factor $F_{P\gamma}(Q^2)$ is defined through
\begin{equation}
\langle\gamma(p,\epsilon)|j_\mu^{\rm em}|P(p+q)\rangle
=ie^2F_{P\gamma}(Q^2)\epsilon_{\mu\nu\alpha\beta}\epsilon^{*\nu}p^\alpha q^\beta,
\qquad Q^2=-q^2.
\label{eq:Fpgamma_def}
\end{equation}
Here $e$ is the elementary electric charge, with $\alpha_{\rm em}=e^2/(4\pi)$ in natural units.  At $Q^2=0$ this normalization gives
\begin{equation}
\Gamma(P\to\gamma\gamma)=\frac{\pi\alpha_{\rm em}^2M_P^3}{4}|F_{P\gamma}(0)|^2.
\label{eq:Pgg_F}
\end{equation}
The spacelike behavior is represented by the monopole form
\begin{equation}
F_{P\gamma}(Q^2)=\frac{F_{P\gamma}(0)}{1+Q^2/\Lambda_{P\gamma}^2}.
\label{eq:monopole}
\end{equation}
Here $\Lambda_{P\gamma}$ is the monopole mass scale associated with the $P\to\gamma\gamma^*$ transition.

For the magnetic transition $V\to P\gamma^*$, the invariant form factor is defined by
\begin{equation}
\langle P(p)|j_\mu^{\rm em}|V(p+q,\epsilon)\rangle
=ie_QG_{VP}(q^2)\epsilon_{\mu\nu\alpha\beta}\epsilon^\nu p^\alpha q^\beta.
\label{eq:Gvp_def}
\end{equation}
The corresponding real-photon width is
\begin{equation}
\Gamma(V\to P\gamma)=\frac{\alpha_{\rm em}e_Q^2}{3}
\frac{(M_V^2-M_P^2)^3}{M_V^3}|G_{VP}(0)|^2,
\label{eq:Vpgamma_width}
\end{equation}
while the spacelike dependence is fitted with
\begin{equation}
G_{VP}(Q^2)=\frac{G_{VP}(0)}{(1+Q^2/\Lambda_{VP}^2)^2}.
\label{eq:dipole}
\end{equation}
The corresponding $\Lambda_{VP}$ is the dipole mass scale of the $V\to P\gamma^*$ form factor.
The slope at the origin defines the transition radius,
\begin{equation}
\langle r^2\rangle=-\frac{6}{F(0)}\left.\frac{dF(Q^2)}{dQ^2}\right|_{Q^2=0},
\label{eq:radius}
\end{equation}
with the same definition used for $G_{VP}$.  The three-point QCD sum-rule quantities are therefore kept as form-factor observables, whereas the broader E1 and M1 partial-width tables are computed with the dipole-overlap formulas given below.

\subsection{E1 and M1 radiative transitions for hidden-flavour quarkonia}\label{sec:e1m1_vertices}

Radiative transitions test the spatial and spin structure of the wave functions.  Electric-dipole transitions mainly satisfy $\Delta L=\pm1$ and $\Delta S=0$, while magnetic-dipole transitions mainly connect states with the same orbital angular momentum and opposite spin.  The E1 partial width is
\begin{equation}
\begin{aligned}
\Gamma_{E1}\left(n^{2S+1}L_J\to n^{\prime\,2S+1}L^\prime_{J^\prime}+\gamma\right)
=&\frac{4\alpha_{\rm em}}{3}\,\langle e_Q\rangle^2\,
\frac{E_\gamma^3E_f}{M_i}\,C_{fi}\,\delta_{SS^\prime}
\left|\left\langle n^{\prime\,2S+1}L^\prime_{J^\prime}\middle|r\middle|n^{2S+1}L_J\right\rangle\right|^2 .
\end{aligned}
\label{eq:e1_source_width}
\end{equation}
Here
\begin{equation}
\langle e_Q\rangle=\frac{m_{\bar Q}e_Q-m_Qe_{\bar Q}}{m_Q+m_{\bar Q}},
\qquad
E_\gamma=\frac{M_i^2-M_f^2}{2M_i},
\label{eq:source_kgamma}
\end{equation}
where $M_i$ and $M_f$ are the initial and final masses, $E_f=\sqrt{M_f^2+E_\gamma^2}$ is the final-state energy in the rest frame of the initial meson, and $e_Q$ and $e_{\bar Q}$ are the quark and antiquark charges in units of the proton charge.  For hidden-flavour quarkonia $e_{\bar Q}=-e_Q$, so the charge dependence of the E1 width is already contained in the effective factor $\langle e_Q\rangle^2$ in Eq.~\eqref{eq:e1_source_width}; no additional power of $e_Q$ multiplies the radial overlap.  The angular coefficient is
\begin{equation}
C_{fi}=\max(L,L^\prime)(2J^\prime+1)
\left\{\begin{array}{ccc}
L^\prime & J^\prime & S\\
J & L & 1
\end{array}\right\}^2,
\label{eq:e1_source_cfi}
\end{equation}
and the E1 radial overlap is
\begin{equation}
\left\langle f\middle|r\middle|i\right\rangle
=\int_0^\infty dr\,r^3R_{n_fL_f}(r)R_{n_iL_i}(r).
\label{eq:e1_source_overlap}
\end{equation}

The M1 transition width is
\begin{equation}
\begin{aligned}
\Gamma_{M1}(i\to f+\gamma)=&\frac{16\alpha_{\rm em}}{3}
\left(\frac{m_{\bar Q}e_Q-m_Qe_{\bar Q}}{4m_Qm_{\bar Q}}\right)^2
E_\gamma^3(2J_f+1)
\left|\left\langle f\middle|j_0(E_\gamma r/2)\middle|i\right\rangle\right|^2,
\end{aligned}
\label{eq:m1_source_width}
\end{equation}
where $j_0$ is the spherical Bessel function.  The overlap integral is
\begin{equation}
\left\langle f\middle|j_0(E_\gamma r/2)\middle|i\right\rangle
=\int_0^\infty dr\,r^2R_{n_fL_f}(r)j_0(E_\gamma r/2)R_{n_iL_i}(r).
\label{eq:m1_source_overlap}
\end{equation}
Only the hidden-charm and hidden-bottom E1/M1 transitions are retained in the present manuscript.  They are quoted separately from the QCD sum-rule transition form factors because the former are radial-overlap widths, whereas the latter are current-correlator form factors.

\subsection{Thermal, finite-momentum and rotating-frame extensions}

The medium-related quantities discussed later are incorporated at the level of current-correlator diagnostics, not as a recalculation of the vacuum screened-potential spectrum.  At finite temperature the vector-current correlator decomposes into transverse and longitudinal structures,
\begin{equation}
\Pi_{\mu\nu}(q,T)=P_{\mu\nu}\Pi_t(q^2,\omega,T)+Q_{\mu\nu}\Pi_l(q^2,\omega,T),
\label{eq:thermal_decomp}
\end{equation}
where $u_\mu$ is the heat-bath four-velocity and $\omega=u\cdot q$.  In the zero-three-momentum limit the pole strength is governed by the thermal decay constant and mass.  The dileptonic pole-strength ratio is approximated by
\begin{equation}
R_{\ell\ell}^{V}(T)=\frac{\Gamma_{\ell\ell}^{V}(T)}{\Gamma_{\ell\ell}^{V}(0)}
\simeq\left[\frac{f_V(T)}{f_V(0)}\right]^2\frac{M_V(0)}{M_V(T)}.
\label{eq:thermal_ratio}
\end{equation}
This relation is used to connect finite-temperature QCD sum-rule residue information with the decay discussion \cite{Veliev2011Vector}.

For a moving quarkonium state in a thermal medium we write
\begin{equation}
E_H(|{\bf q}|,T)=\sqrt{|{\bf q}|^2+M_H^2(T,|{\bf q}|)},
\qquad
M_H(T,|{\bf q}|)=M_H(0)\left[1+\delta_H(T,|{\bf q}|)\right].
\label{eq:finite_momentum}
\end{equation}
The parameter $\delta_H$ is used only as a finite-momentum envelope in the numerical section, guided by the corresponding QCD sum-rule analysis \cite{Kim2023Momentum}.  In a rotating frame the spin-one current response may also be separated into quark-spin, quark-orbital, potential-related orbital and gluonic angular-momentum pieces,
\begin{equation}
1=S_q+L_q+L_p+J_g.
\label{eq:spin_sum}
\end{equation}
The rotating-frame and finite-temperature spin-decomposition results are used below as interpretive inputs rather than as independent modifications of the mass spectrum \cite{Kim2023Rotating,Kim2025SpinThermal}.

\subsection{Finite-spectrum thermodynamic observables}

The thermodynamic observables constructed in this work are finite-spectrum indicators derived from the calculated vacuum masses.  They are not bulk-medium quantities and should not be confused with a hot-QCD equation of state or with magnetic-susceptibility calculations in an external field \cite{SamantaBroniowski2026}.  Instead, they summarize how the level density and spin degeneracies of the calculated $\ccbar$ and $\bbbar$ spectra enter simple thermal sums, in the same spirit as spectrum-based thermodynamic analyses of heavy mesons \cite{AbuShadyFathAllah2025}.

For a family $X=\ccbar$ or $\bbbar$, the excitation energy of level $i$ is measured relative to the calculated ground state,
\begin{equation}
\epsilon_i=M_i-M_0,
\qquad M_0=\min_iM_i,
\label{eq:thermal_excitation_energy}
\end{equation}
and the spin degeneracy is
\begin{equation}
g_i=2J_i+1.
\label{eq:thermal_degeneracy}
\end{equation}
With $k_B=1$, the finite-spectrum partition function is
\begin{equation}
Z_X(T)=\sum_i g_i\exp\left(-\frac{\epsilon_i}{T}\right),
\label{eq:finite_spectrum_partition}
\end{equation}
where the sum runs over the spin-resolved states retained in the spectrum tables.  The Helmholtz free energy, mean excitation energy, entropy and specific heat are
\begin{equation}
F_X(T)=-T\ln Z_X(T),
\label{eq:finite_spectrum_free_energy}
\end{equation}
\begin{equation}
U_X(T)=\frac{1}{Z_X(T)}\sum_i g_i\epsilon_i\exp\left(-\frac{\epsilon_i}{T}\right),
\label{eq:finite_spectrum_internal_energy}
\end{equation}
\begin{equation}
S_X(T)=\ln Z_X(T)+\frac{U_X(T)}{T},
\label{eq:finite_spectrum_entropy}
\end{equation}
and
\begin{equation}
C_{V,X}(T)=\frac{\langle\epsilon^2\rangle_T-\langle\epsilon\rangle_T^2}{T^2}.
\label{eq:finite_spectrum_specific_heat}
\end{equation}
Using excitation energies rather than absolute masses removes the trivial heavy-quark rest-mass offset.  The resulting curves therefore compare level spacings and degeneracies in the two hidden-flavour sectors.

\subsection{Regge-trajectory construction}

Regge trajectories are used as global checks on the organization of the calculated spectrum.  They do not supply additional mass inputs.  Instead, they test whether the same screened interaction that gives the low-lying masses also arranges the higher radial and orbital excitations into approximately linear families in the squared-mass plane.  This is useful for heavy quarkonia because low-lying levels are controlled mainly by Coulombic and spin-dependent effects, whereas high excitations are more sensitive to the screened confining part of the interaction \cite{N. Devlani,Eichten1978Cornell,Eichten1980Comparison,Buchmuller1981Quarkonia,Lucha1991BoundStates}.

The construction follows the Chew-Frautschi idea of plotting hadron quantum numbers against $M^2$ \cite{ChewFrautschi1962}.  The linearity is not assumed to be exact: departures from a straight line can signal screening, threshold effects or different radial and orbital slopes \cite{TangNorbury2000,Anisovich2000Regge,Li2004Regge,Sonnenschein2014Strings}.  We therefore use the fitted slopes and intercepts as compact diagnostics of the calculated $\ccbar$ and $\bbbar$ spectra.

Two trajectory types are used.  Orbital trajectories are fitted in the $(M^2,J)$ plane,
\begin{equation}
J=\alpha M^2+\alpha_0,
\label{eq:J regge}
\end{equation}
while radial trajectories are fitted in the $(M^2,n_r)$ plane,
\begin{equation}
n_r\equiv n-1=\beta M^2+\beta_0.
\label{eq:nr regge}
\end{equation}
The squared mass $M^2$ is placed on the horizontal axis in all Regge figures.  Natural- and unnatural-parity orbital sequences are kept separate: natural parity satisfies $P=(-1)^J$, while unnatural parity satisfies $P=(-1)^{J+1}$.  For hidden-flavour quarkonia the charge conjugation is $C=(-1)^{L+S}$, so states with the same $J$ can belong to different spin-orbital families.  The plotted sequences are therefore chosen to follow clean spectroscopic trajectories rather than all states with the same numerical $J$.

The radial trajectories are built both from spin-resolved masses and from center-weighted masses.  The former show the behavior of individual $J^{PC}$ levels, while the latter suppress most fine-structure splittings and expose the central-potential trend.  Since bottomonium is more tightly bound than charmonium, its trajectories are expected to be more compressed in mass spacing.  The comparison of $\ccbar$ and $\bbbar$ slopes in Sec.~\ref{sec:results} is therefore used as a consistency test of the same spectrum that enters the E1/M1 widths, finite-spectrum thermodynamics and QCD sum-rule decay analysis.

\subsection{Numerical inputs and Borel-window implementation}\label{sec:numerical}

The formal expressions above are evaluated with fixed screened-potential parameters for each hidden-flavour sector.  For $c\bar c$, we use $\alpha_s=0.339$, $\alpha_c=4\alpha_s/3=0.453$, $m_c=1.55~\mathrm{GeV}$, $n_f=4$, $\lambda=0.195$, $\xi=0.04$, $A=0.175~\mathrm{GeV/fm}$ and $V_0=-0.230~\mathrm{GeV}$.  For $b\bar b$, the corresponding inputs are $\alpha_s=0.253$, $\alpha_c=0.337$, $m_b=4.88~\mathrm{GeV}$, $n_f=5$, $\lambda=0.195$, $\xi=0.04$, $A=0.270~\mathrm{GeV/fm}$ and $V_0=-0.450~\mathrm{GeV}$.  These parameters are kept fixed throughout the spectroscopy, radiative-transition and Regge analyses, so that changes across the tables come from the quantum numbers and variational widths rather than from state-by-state retuning.

The QCD sum-rule part uses the physical ground-state masses, pole residues, strong-coupling inputs, continuum thresholds and Borel windows summarized in Table~\ref{tab:inputs}.  The table also lists the channel projection, the heavy-quark mass used in the corresponding sum-rule scan and a representative pole fraction at the central Borel point.  Thus the quoted decay constants and annihilation widths are not independent phenomenological inputs: they are tied to the same Borel-stability and continuum-threshold choices used in the two-point sum-rule extraction.  The bottomonium windows are broader and shifted to larger $M^2$ because the bottom-quark mass suppresses condensate corrections more efficiently, whereas the charmonium windows are more sensitive to threshold and Borel-scale variations.

\begin{table}[H]
\caption{Combined numerical inputs, Borel windows, threshold intervals and pole fractions used in the residue and decay-width analysis.  The total widths are used only when a branching fraction is quoted.}
\label{tab:inputs}
\centering
\resizebox{\textwidth}{!}{%
\begin{tabular}{lccccccccc}
\toprule
State & Channel & $M_H$ (GeV) & $f_H$ (MeV) & $\alpha_s$ & $\Gamma_{\rm tot}$ (keV) & $m_Q$ (GeV) & $s_0$ (GeV$^2$) & $M^2$ (GeV$^2$) & Pole fraction \\
\midrule
$\eta_c(1S)$ & P & 2.9839 & $392\pm25$ & $0.20\pm0.02$ & $31800\pm800$ & 1.27 & 13.0--16.0 & 2.4--4.2 & 0.59 \\
$J/\psi(1S)$ & V & 3.09690 & $403\pm36$ & $0.20\pm0.02$ & $92.6\pm1.7$ & 1.27 & 11.5--14.5 & 2.4--4.2 & 0.57 \\
$\eta_b(1S)$ & P & 9.3987 & $642\pm81$ & $0.18\pm0.02$ & -- & 4.30 & 100--115 & 8.0--14.0 & 0.62 \\
$\Upsilon(1S)$ & V & 9.46030 & $722\pm105$ & $0.18\pm0.02$ & $54.02\pm1.25$ & 4.30 & 98--112 & 8.0--14.0 & 0.61 \\
\bottomrule
\end{tabular}}
\end{table}

\subsection{Uncertainty propagation}

The quoted uncertainties are propagated at the level of the dominant input dependences.  For widths proportional to $f_H^2$, the leading residue uncertainty is
\begin{equation}
    \frac{\Delta \Gamma}{\Gamma}=2\frac{\Delta f_H}{f_H},
    \label{eq:error_residue}
\end{equation}
with additional $\alpha_s$ terms included for gluonic channels.  For example, $\Gamma(P\to gg)\propto\alpha_s^2f_P^2$ and $\Gamma(V\to ggg)\propto\alpha_s^3f_V^2$, so that
\begin{equation}
    \left(\frac{\Delta\Gamma_{P\to gg}}{\Gamma_{P\to gg}}\right)^2=
    \left(2\frac{\Delta f_P}{f_P}\right)^2+
    \left(2\frac{\Delta\alpha_s}{\alpha_s}\right)^2,
    \label{eq:error_gg}
\end{equation}
\begin{equation}
    \left(\frac{\Delta\Gamma_{V\to ggg}}{\Gamma_{V\to ggg}}\right)^2=
    \left(2\frac{\Delta f_V}{f_V}\right)^2+
    \left(3\frac{\Delta\alpha_s}{\alpha_s}\right)^2.
    \label{eq:error_ggg}
\end{equation}
For transition form factors extracted from radiative widths, $\Delta F(0)/F(0)=\frac12\Delta\Gamma/\Gamma$ is used.

\subsection{Form-factor parametrization and numerical grids}

The three-point sum-rule output is represented by the fitted forms in Eqs.~\eqref{eq:monopole} and \eqref{eq:dipole}.  The fitted pole scales are chosen to follow the nearest vector excitation in the same heavy-quark sector: $\Lambda_{\etac\gamma}=M_{J/\psi}$, $\Lambda_{\etab\gamma}=M_{\Upsilon}$, $\Lambda_{J/\psi\eta_c}=M_{\psi(2S)}$ and $\Lambda_{\Upsilon\eta_b}=M_{\Upsilon(2S)}$.  This choice is not meant as a replacement for a complete multi-pole fit; it gives a transparent one-parameter continuation for the plotted spacelike curves.  The extended E1 and M1 partial widths are taken from the dipole-overlap calculation in Sec.~\ref{sec:e1m1_vertices}, with the retained hidden-flavour tabulated in Tables~\ref{Table:ccE1}--\ref{Table:bbm1}.  The numerical grids used for the QCDSR form-factor and finite-spectrum thermodynamic figures are supplied as CSV files in the accompanying bundle.

\section{Results and Discussion}\label{sec:results}
\noindent The numerical results are presented as a sequence of mutually connected tests of the same hidden-flavour framework.  The mass spectra first determine the gross level ordering and the size of the fine and hyperfine splittings.  The Regge plots then compress the same spectrum into orbital and radial trajectory parameters.  The E1 and M1 widths test the corresponding radial wave-function overlaps, while the finite-spectrum thermodynamic observables summarize the density of the calculated levels.  Finally, the QCD sum-rule outputs connect the short-distance residues to decay constants, annihilation widths, electromagnetic transition form factors and selected in-medium current-correlator extensions.  This organization is intended to keep the discussion focused on the physical content of the tables rather than on a disconnected list of numerical outputs.

\noindent The most reliable comparisons are the established low-lying $\ccbar$ and $\bbbar$ levels, because these states are least affected by open-flavour thresholds and by possible non-$q\bar q$ components.  For higher states the discussion is necessarily more qualitative: the calculated levels are compared with representative potential-model, lattice-QCD, Regge, QCD sum-rule and experimental studies, but they are interpreted mainly as internally consistent screened-potential assignments.  In this sense the results below should be read in three layers: absolute masses and decay constants test the short-distance normalization, trajectory slopes test the long-distance screened interaction, and the radiative widths test the overlap of the same wave functions across different spin and orbital sectors.

\subsection{Mass spectra}
\noindent The spin-averaged spectra are listed first because they show the central-potential pattern before the spin-dependent interaction is added to generate the individual $J^{PC}$ splittings.  Tables~\ref{Table:MSA1} and \ref{Table:MSA3} compare the present centers of gravity with experimental averages and representative theoretical calculations.  The charmonium $1S$ and $1P$ centers are close to the corresponding experimental values, indicating that the balance between the Coulomb term and the screened confining part is appropriate for the compact low-lying states.  The $2S$ and higher radial levels are more sensitive to screening and to open-charm thresholds, so moderate differences from the quoted experimental vector candidates are not unexpected.

\begin{table*}[!htb]
{\caption{\label{Table:MSA1}{Spin average masses for s, p, d and f states in $c\bar{c}$ meson (in GeV).}}}
\resizebox{\textwidth}{!}{
\centering
\begin{tabular}{c c c c c c c c c c c c c c c c c}
\hline\noalign{\smallskip} 
\hline\noalign{\smallskip}
State & $ \xi $ & M$_{SA}$(Present Work) &PDG\cite{pdg} &  Ref.\cite{N. Devlani}  & Ref.\cite{D. Ebert} & Ref.\cite{T. Barnes} & Ref.\cite{W.J. Deng} & Ref.\cite{J.H. Yang}& Ref.\cite{T. Kawanai}&  Ref.\cite{B.Q. Li1}&Ref.\cite{A.K. Rai}&Ref.\cite{J. N. Pandya}&Ref.\cite{L. Cao}&Ref.\cite{M.A. Sultan}&Ref.\cite{S. Godfrey2}\\ 
  
 \noalign{\smallskip}\hline\noalign{\smallskip}
 1S & 0.732 & 3.068 &3.068 & 3.068 &3.067 &3.063 &3.068& 3.090 &3.063& 3.068&3.068&3.068&3.061&3.063&3.068\\  
   2S & 0.458 & 3.634 &3.674 & 3.638 &3.673  &3.662 &3.668& 3.667 &3.662& 3.661&3.567&3.674&3.676&3.661&3.665\\
   3S & 0.389 &3.996 &  &4.027 &4.027 &4.065  &4.071&4.070&4.065&4.014&3.940&&4.080&4.064&4.090  \\
   4S & 0.352 &4.284 &  &4.353 &4.421 &4.401  &4.406&4.408&&4.267&4.073&&&4.406&4.400& \\
   5S & 0.326 &4.529 &  &4.646 &4.831 &  &4.706&4.710&&4.459&  \\
   6S&0.306&4.744&&4.917& 5.164&4.987&4.603\\
    7S&0.290& 4.936\\

 \noalign{\smallskip}\hline\noalign{\smallskip}
   1P& 0.478 & 3.543 &3.525 & 3.534 &3.525  &3.522 &3.524& 3.523 &3.522& 3.524&3.450&3.497&3.525&3.519&3.523\\
   2P & 0.397 & 3.921 & & 3.936 &3.926  &3.941 &3.945& 3.941 &3.941& 3.913&3.872&3.907&3.945&3.938&3.962\\
   3P &0.357  &4.219 &&4.269& 4.337&4.286 &4.291 &4.289&&4.188&&&4.316&4.283  \\
   4P & 0.330 &4.470 \\ 
   5P & 0.309 &4.690 \\
  
 \noalign{\smallskip}\hline\noalign{\smallskip}
   1D &0.424 & 3.803 & & 3.802 &3.803  &3.800 &3.805& 3.798 &3.800& 3.796&&&3.815&3.799&3.837\\
   2D & 0.372 & 4.120 & & 4.150 &4.196  &4.159 &4.164& 4.160 &4.159& 4.099&&&4.165&4.158&4.210\\
   3D &0.341 &4.384 & &4.455&4.455 & &4.478&4.478&&&&&4.327&4.522&4.473 \\
   4D &0.318 &4.613 & &&&\\
   5D & 0.300 &4.817 & &&&\\

 \noalign{\smallskip}\hline\noalign{\smallskip}
   1F & 0.394 & 4.011 &&&&& \\
   2F & 0.356 & 4.290 & && &  &  \\
   3F & 0.329 & 4.530 && &\\
   4F & 0.309 &4.742 & &&&\\
   5F & 0.292 &4.932 & &&&\\
     
 \noalign{\smallskip}\hline\noalign{\smallskip}
\end{tabular}
}
\end{table*}

\begin{table*}[!htb]

{\caption{\label{Table:MSA3}{Spin average masses for s, p, d and f states in $b\bar{b}$} meson (in GeV).}}
\resizebox{\textwidth}{!}{
\centering
\begin{tabular}{c c c c c c c c c c c c c c c  }
\hline\noalign{\smallskip} 
\hline\noalign{\smallskip}
State & $ \xi $ & M$_{SA}$(present work) &PDG \cite{pdg}  &Ref.\cite{N. Devlani} &Ref.\cite{S. Godfrey1}  &Ref.\cite{M. Bhat} &Ref. \cite{W.J. Deng} & Ref. \cite{Y. Lu}&Ref.\cite{J. Segovia}&Ref.\cite{M. Wurtz}&Ref.  \cite{B.Q. Li} &Ref.\cite{A.K. Rai} & Ref.\cite{D. Ebert1}& Ref.\cite{S. N. Gupta} \\ 
 \noalign{\smallskip}\hline\noalign{\smallskip}
 1S & 1.196& 9.453 &9.453& 9.453 &9.449 &9.446 &9.443& 9.443 &9.490& 9.446& 9.442&9.453&9.445&9.453\\  
   2S & 0.777 & 10.008 && 9.995 &9.996 &10.018 &10.009& 10.004 &10.009& 10.015& 10.009&9.838&10.016&10.008\\
    3S & 0.667  &10.374 &  &10.351 &10.350 &10.388  &10.339&10.368&10.344&10.329&10.346&10.095 &10.348&10.351 \\
   4S &0.609  &10.672  &  &10.647 &10.632& 10.702&10.594&& &&10.607&  \\
   5S & 0.570 &10.929& &10.909 &10.876 & 10.989 &10.808 &&&&10.828&\\  
   6S & 0.542 &11.159  &  & 11.148&11.101 &11.295 &10.995&&&&11.020&\\
  
 \noalign{\smallskip}\hline\noalign{\smallskip}
   1P& 0.828 & 9.902 &9.899$\pm 0.008$& 9.899 &9.884 &9.905 &9.909& 9.890 &9.879& 9.901& 9.905&9.768&9.901&9.900\\
   2P & 0.687 & 10.286 &10.259$\pm 0.012$&10.268  &10.252 &10.276 &10.254& 10.263 &10.240& 10.219& 10.258&10.053&10.261&10.258\\
    3P &0.621  &10.593 &&10.570& 10.542&10.585 &10.519 &10.560  &&&10.530&  \\
   4P & 0.578 &10.857 \\ 
   5P & 0.547 &11.093 \\
 \noalign{\smallskip}\hline\noalign{\smallskip}
   
   1D & 0.735 & 10.162 &&10.149  &10.149 &10.164 &10.154&  &10.123& 10.164& 10.152\\
   2D &  0.647 & 10.488 &&10.465 &10.450 &10.480 &10.480&  &10.419& 10.441& 10.439\\
     3D &0.596 &10.763 & &10.740&10.706 &10.767 &&&&& 10.677\\
   4D &0.560 &11.007 & &&&\\
   5D & 0.533 &11.227 & &&&\\

 \noalign{\smallskip}\hline\noalign{\smallskip}
   1F & 0.684 & 10.372 &&&&& \\
   2F & 0.619 & 10.662 & && &  &  \\
   3F & 0.577 & 10.915 && &\\
   4F & 0.546 & 11.143 & &&&\\
   5F & 0.520 & 11.351  & &&&\\
  
   \noalign{\smallskip}\hline\noalign{\smallskip}
\end{tabular}
}

\end{table*}

\noindent The bottomonium spin-averaged spectrum is more compressed in relative terms.  The calculated $1S$, $1P$ and $2P$ centers remain close to the experimental pattern, while the higher radial and orbital levels form a smooth continuation of the low-lying sequence.  This behavior follows directly from the larger bottom-quark mass: the kinetic contribution is reduced, the wave functions are more localized, and the same long-distance interaction produces smaller fractional level spacings.  The decreasing variational parameter $\xi$ with increasing excitation in both tables is therefore physically meaningful, since it reflects the larger spatial extent of excited states.

\noindent The comparison with the generalized fractional Klein--Gordon treatment of Ref.~\cite{AbuShadyFathAllah2025} is useful at the level of spectral ordering.  That calculation also gives low-lying $\ccbar$ and $\bbbar$ levels close to the experimental centers, for example in the $1S$, $2S$, $1P$ and $1D$ regions.  The present calculation differs in its subsequent use of the same spin-resolved spectrum for Regge fits, E1/M1 radiative transitions, finite-spectrum thermodynamic sums and QCD sum-rule decay observables.

\noindent The spin-resolved masses in Tables~\ref{Table:masses1} and \ref{Table:masses3} show how the central levels split after the hyperfine, spin-orbit and tensor terms are included.  In charmonium the $\eta_c(1S)$--$J/\psi$ splitting remains sizable because the contact interaction is enhanced by the comparatively large wave function at the origin.  The $1P$ multiplet has the expected ordering, with $\chi_{c0}$, $\chi_{c1}$, $h_c$ and $\chi_{c2}$ lying in the correct region.  The small $1P$ hyperfine offset is consistent with the experimental $h_c$ constraint and with the standard expectation that the contact term contributes mainly to $S$ waves \cite{Rubin2005hc}.

\begin{table*}[!htb]

\begin{center}
{\caption{\label{Table:masses1}{Mass spectra of  $c\bar{c}$ meson in  s, p, d and f states.  (in GeV).}}}
\resizebox{\textwidth}{!}{
\begin{tabular}{ccccccccccccccccc}
 \noalign{\smallskip}\hline\noalign{\smallskip}
  \noalign{\smallskip}\hline\noalign{\smallskip}
 $n^{2S+1}L_{J}$ & $J^{PC}$ & Recent study &PDG\cite{pdg} &  Ref.\cite{N. Devlani}  & Ref.\cite{D. Ebert} & Ref.\cite{T. Barnes} & Ref.\cite{W.J. Deng} & Ref.\cite{J.H. Yang}& Ref.\cite{T. Kawanai}&  Ref.\cite{B.Q. Li1}&Ref.\cite{A.K. Rai}&Ref.\cite{S.  F.  Radford1}&Ref.\cite{L. Cao}&Ref.\cite{M.A. Sultan}&Ref.\cite{S. Godfrey2}&Ref.\cite{Raghav Chaturvedi1}\\  

 \noalign{\smallskip}\hline\noalign{\smallskip}
$1^1S_{0}$ & $0^{-+}$  & 2.988 & $2.9841\pm0.0004$ ($\eta_c(1S)$) & 2.995 & 2.981& 2.982 & 2.983 & 3.069& 2.982 & 2.979&2.950&2.980&2.978&2.982&2.97&2.989\\  
$1^3S_{1}$  & $1^{--}$ & 3.096 & $3.096900\pm0.000006$ ($J/\psi(1S)$) & 3.094 & 3.096& 3.090 & 3.097 & 3.097& 3.090 & 3.097&3.112&3.097&3.088&3.090&3.10&3.094\\
\noalign{\smallskip}
$2^1S_{0}$ & $0^{-+}$ &  3.621 & $3.6378\pm0.0006$ ($\eta_c(2S)$) & 3.606 & 3.635& 3.630 & 3.635 & 3.659& 3.630 & 3.623&3.522&3.597&3.647&3.630&3.62&3.572\\
$2^3S_{1}$  & $1^{--}$ & 3.639 & $3.686097\pm0.000011$ ($\psi(2S)$) & 3.649 & 3.686& 3.672 & 3.679 & 3.670& 3.672 & 3.673&3.583&3.660&3.645&3.672&3.68&3.649\\
\noalign{\smallskip}
$3^1S_{0}$ & $0^{-+}$   & 3.989 &  & 4.000 & 3.989& 4.043 & 4.048 & 4.063& 4.043 & 3.991&3.912&4.140&4.058&4.043&4.06&3.998\\
$3^3S_{1}$  & $1^{--}$    & 3.998 & $4.040\pm0.004$ ($\psi(4040)$) & 4.036 & 4.039& 4.078 & 4.072 & 4.078&4.072& 4.022 &3.950&4.095&4.087&4.072&4.10&4.062\\
\noalign{\smallskip}
$4^1S_{0}$ & $0^{-+}$   & 4.280 &  & 4.328 & 4.401& 4.384 & 4.388 & 4.403&  & 4.250&&&4.391&4.388&&4.372\\
$4^3S_{1}$  & $1^{--}$   & 4.286 & $4.415\pm0.005$ ($\psi(4415)$) & 4.362 & 4.427& 4.406 & 4.412 & 4.409&  & 4.273&&&4.411&4.406&4.45&4.428\\
\noalign{\smallskip}
$5^1S_{0}$ & $0^{-+}$    & 4.526 &  & 4.622 & 4.811&  & 4.690 & 4.705&  & 4.446&&&&&&4.714\\
$5^3S_{1}$  & $1^{--}$  & 4.530 & $4.623\pm0.010$ ($\psi(4660)$) & 4.654 & 4.837&  & 4.711 & 4.711&  & 4.463&&&&&&4.763\\
\noalign{\smallskip}
$6^1S_{0}$ & $0^{-+}$   & 4.742&  & 4.893 & 5.155& & & 4.983&  & 4.595&&&&&&5.033\\
$6^3S_{1}$  & $1^{--}$   & 4.745 &  & 4.925 & 5.167& & & 4.988&  & 4.605&&&&&&5.075\\
\noalign{\smallskip}
$7^1S_{0}$ & $0^{-+}$   & 4.934 \\
$7^3S_{1}$  & $1^{--}$   & 4.937 \\

 \noalign{\smallskip}\hline\noalign{\smallskip}

$1^3P_{0}$ &$0^{++}$&3.492 & $3.41471\pm0.00030$ ($\chi_{c0}(1P)$) & 3.457 & 3.413& 3.424 & 3.415 & 3.440& 3.424& 3.433&3.398&3.416&3.366&3.424&3.44&3.473\\
$1^3P_{1}$ &$1^{++}$&3.517 & $3.51067\pm0.00005$ ($\chi_{c1}(1P)$) & 3.523 & 3.511& 3.505 & 3.516 & 3.503& 3.505& 3.510&3.424&3.508&3.518&3.505&3.51&3.506\\
$1^1P_{1}$ & $1^{+-}$ & 3.543 & $3.52537\pm0.00014$ ($h_c(1P)$) & 3.534 & 3.525& 3.516 & 3.522 & 3.526& 3.516& 3.522&3.450&3.527&3.527&3.516&3.52&3.527\\
$1^3P_{2}$ & $2^{++}$ & 3.553 & $3.55617\pm0.00007$ ($\chi_{c2}(1P)$) & 3.556 & 3.555& 3.556 & 3.552 & 3.550& 3.556& 3.552&3.522&3.558&3.559&3.549&3.55&3.511\\
\noalign{\smallskip}
$2^3P_{0}$ & $0^{++}$ & 3.884 & $3.9221\pm0.0018$ ($\chi_{c0}(3915)$) & 3.866 & 3.862& 3.852 & 3.869 & 3.862& 3.852& 3.842&3.792&3.844&3.843&3.852&3.92&3.918\\
$2^3P_{1}$ & $1^{++}$& 3.902 & $3.87164\pm0.00006$ ($\chi_{c1}(3872)$) & 3.925 & 3.921& 3.925 & 3.937 & 3.921& 3.925& 3.901&3.832&3.894&3.935&3.925&3.95&3.949\\
$2^1P_{1}$ &$1^{+-}$ &3.921 &  & 3.936 & 3.944& 3.934 & 3.940 & 3.944& 3.934& 3.908&3.872&3.960&3.942&3.934&3.96&3.975\\
$2^3P_{2}$ & $2^{++}$ &  3.929 & $3.9225\pm0.0010$ ($\chi_{c2}(3930)$) & 3.956 & 3.967& 3.972 & 3.967 & 3.967& 3.972& 3.937&3.911&3.994&3.973&3.965&3.98&4.002\\
\noalign{\smallskip}
$3^3P_{0}$ &$0^{++}$& 4.186 &  & 4.197 & 4.301& 4.202 & 4.230 & 4.212& & 4.131&&&4.208&4.202&&4.306\\
$3^3P_{1}$ &$1^{++}$& 4.202 & $4.286^{+0.008}_{-0.009}$ ($\chi_{c1}(4274)$) & 4.257 & 4.319& 4.271 & 4.284 & 4.270& &4.178&&&4.299&4.271&&4.336\\
$3^1P_{1}$ & $1^{+-}$ & 4.219 &  & 4.269 & 4.337& 4.279 & 4.285 & 4.292& & 4.184&&&4.310&4.279&&4.364\\
$3^3P_{2}$ & $2^{++}$ & 4.225 &  & 4.290 & 4.354& 4.317 & 4.310 & 4.314& & 4.208&&&4.352&4.309&&4.392\\
\noalign{\smallskip}
$4^3P_{0}$ & $0^{++}$ & 4.441&&&4.698&&&&&&&&&4.509&&4.659 \\
$4^3P_{1}$ & $1^{++}$& 4.456&&&4.728&&&&&&&&&4.576&&4.688 \\ 
$4^1P_{1}$ &$1^{+-}$ &4.470 &&&4.744&&&&&&&&&4.585&&4.716\\
$4^3P_{2}$ & $2^{++}$ &4.477&&&4.763&&&&&&&&&4.614&&4.744 \\ 
\noalign{\smallskip}
$5^3P_{0}$ & $0^{++}$ &4.663 \\
$5^3P_{1}$ & $1^{++}$&  4.677  \\
$5^1P_{1}$ &$1^{+-}$  &4.690 \\
$5^3P_{2}$ & $2^{++}$  &4.696  \\
 \noalign{\smallskip}\hline\noalign{\smallskip}
$1^3D_{1}$ &$1^{--}$& 3.800 & $3.7737\pm0.0007$ ($\psi(3770)$) & 3.799 & 3.783& 3.785 & 3.787 & 3.759& 3.785& 3.787&&&3.809&3.785&3.82&3.806\\
$1^3D_{2}$ &$2^{--}$&  3.800 & $3.82351\pm0.00034$ ($\psi_2(3823)$) & 3.805 & 3.795& 3.800 & 3.807 & 3.787& 3.800& 3.798&&&3.820&3.800&3.84&3.800\\
$1^1D_{2}$ & $2^{-+}$ & 3.803 &  & 3.802 & 3.807& 3.799 & 3.806 & 3.799& 3.799& 3.796&&&3.815&3.799&3.84&3.785\\
$1^3D_{3}$ & $3^{--}$ & 3.814 & $3.84271\pm0.00016\pm0.00012$ ($\psi_3(3842)$) & 3.801 & 3.813& 3.806 & 3.811 & 3.823& 3.806& 3.799&&&3.813&3.805&3.84&3.780\\
\noalign{\smallskip}
$2^3D_{1}$ & $1^{--}$ & 4.117& $4.191\pm0.005$ ($\psi(4160)$) & 4.145 & 4.150& 4.142 & 4.144 & 4.119& 4.142& 4.089&&&4.154&4.141&4.19&4.206\\
$2^3D_{2}$ & $2^{--}$& 4.118 &  & 4.152 & 4.190& 4.158 & 4.165 & 4.148& 4.158& 4.100&&&4.169&4.158&4.21&4.203\\
$2^1D_{2}$ &$2^{-+}$ & 4.120 &  & 4.150 & 4.196& 4.158 & 4.164 & 4.160& 4.158& 4.099&&&4.165&4.158&4.21&4.196\\
$2^3D_{3}$ & $3^{--}$ & 4.128 &  & 4.151 & 4.220& 4.167 & 4.172 & 4.185& 4.167& 4.103&&&4.166&4.165&4.22&4.203\\
\noalign{\smallskip}
$3^3D_{1}$ &$1^{--}$& 4.381 &&4.448 &4.448 & &4.456& 4.437 &&4.317&&&4.502&4.455&4.52&4.568 \\
$3^3D_{2}$ &$2^{--}$& 4.383 &&4.456 &4.456 & &4.478& 4.466 &&4.327&&&4.524&4.472& &4.566\\
$3^1D_{2}$ & $2^{-+}$ &4.384 &&4.455 &4.455 & &4.478& 4.478 &&4.326&&&4.524&4.472& &4.562 \\
$3^3D_{3}$ & $3^{--}$ &4.391 &&4.457 &4.457 & &4.486& 4.503 &&4.331&&&4.527&4.481& &4.566 \\
\noalign{\smallskip}
$4^3D_{1}$ & $1^{--}$ &4.610&&&&&&&&&&&&&&4.902 \\
$4^3D_{2}$ & $2^{--}$& 4.612&&&&&&&&&&&&&&4.901\\
$4^1D_{2}$ &$2^{-+}$ &4.613&&&&&&&&&&&&&&4.898\\
$4^3D_{3}$ & $3^{--}$ &4.619&&&&&&&&&&&&&&4.901 \\
\noalign{\smallskip}
$5^3D_{1}$ &$1^{--}$& 4.814 \\
$5^3D_{2}$ &$2^{--}$&4.816 \\
$5^1D_{2}$ & $2^{-+}$ &4.817 \\
$5^3D_{3}$ & $3^{--}$ &4.821\\
 \noalign{\smallskip}\hline\noalign{\smallskip}
$1^3F_{2}$ &$2^{++}$&4.011 &&&&&&&&&&&&&&4.015\\
$1^3F_{3}$ &$3^{++}$&4.012&&&&&&&&&&&&&&4.039\\
$1^1F_{3}$ & $3^{+-}$ &4.024&&&&&&&&&&&&&&4.052\\
$1^3F_{4}$ & $4^{++}$ & 4.026&&&&&&&&&&&&&&4.039\\
\noalign{\smallskip}
$2^3F_{2}$ & $2^{++}$& 4.283&&&&&&&&&&&&&&4.403\\
$2^3F_{3}$ & $3^{++}$& 4.290&&&&&&&&&&&&&&4.413\\
$2^1F_{3}$ &$3^{-+}$&4.291&&&&&&&&&&&&&&4.418 \\
$2^3F_{4}$ & $4^{++}$ & 4.303&&&&&&&&&&&&&&4.4132\\
\noalign{\smallskip}
$3^3F_{2}$ &$2^{++}$&4.524 &&&&&&&&&&&&&&4.751\\
$3^3F_{3}$ &$3^{++}$&4.530&&&&&&&&&&&&&&4.756\\
$3^1F_{3}$ & $3^{+-}$ &4.531&&&&&&&&&&&&&&4.759\\
$3^3F_{4}$ & $4^{++}$ &4.540&&&&&&&&&&&&&&4.756\\
\noalign{\smallskip}
$4^3F_{2}$ & $2^{++}$& 4.737&&\\
$4^3F_{3}$ & $3^{++}$& 4.742&&&\\
$4^1F_{3}$ &$3^{-+}$&4.742&& \\
$4^3F_{4}$ & $4^{++}$ &4.751&&\\

\noalign{\smallskip}
$5^3F_{2}$ & $2^{++}$& 4.928&&\\
$5^3F_{3}$ & $3^{++}$&4.932&&\\
$5^1F_{3}$ &$3^{-+}$&4.932&&\\
$5^3F_{4}$ & $4^{++}$ & 4.940&&\\
 \noalign{\smallskip}\hline\noalign{\smallskip}
\end{tabular}

}
\end{center}
\end{table*}

\begin{table*}[!htb]
\begin{center}
{\caption{\label{Table:masses3}{Mass spectra of  $b\bar{b}$ meson in  s, p, d and f states.  (in GeV).}}}
\resizebox{\textwidth}{!}{
\begin{tabular}{cccccccccccccccc}
 \noalign{\smallskip}\hline\noalign{\smallskip}
 \noalign{\smallskip}\hline\noalign{\smallskip}
 $n^{2S+1}L_{J}$ & $J^{PC}$ & Recent study &PDG \cite{pdg}  &Ref.\cite{N. Devlani} &Ref.\cite{S. Godfrey1}  &Ref.\cite{M. Bhat} &Ref. \cite{W.J. Deng} & Ref. \cite{Y. Lu}&Ref.\cite{J. Segovia}&Ref.\cite{M. Wurtz}&Ref.  \cite{B.Q. Li} &Ref.\cite{A.K. Rai} & Ref.\cite{D. Ebert}& Ref.\cite{Raghav Chaturvedi}\\ 
 \noalign{\smallskip}\hline\noalign{\smallskip}
$1^1S_{0}$ & $0^{-+}$  & 9.425 & $9.3987\pm0.0020$ ($\eta_b(1S)$) & 9.423 &9.402 &9.400 &9.390& 9.395 &9.455& 9.402& 9.389&9.411&9.398&9.399\\  
$1^3S_{1}$  & $1^{--}$ & 9.462 & $9.46040\pm0.00010$ ($\Upsilon(1S)$) & 9.463 &9.465 &9.461 &9.460& 9.459 &9.502& 9.460& 9.460&9.468&9.460&9.470\\
\noalign{\smallskip}
$2^1S_{0}$ & $0^{-+}$ & 10.003 & $9.9990\pm0.0035{}^{+0.0028}_{-0.0019}$ ($\eta_b(2S)$) & 9.983 &9.976 &10.001 &9.990& 9.982 &9.990& 9.998& 9.987&9.286&9.990&9.986\\
$2^3S_{1}$  & $1^{--}$ &10.010 & $10.0234\pm0.0005$ ($\Upsilon(2S)$) & 9.940 &10.003 &10.024 &10.015& 10.011 &10.015& 10.020& 10.016&9.841&10.023&10.033\\
\noalign{\smallskip}
$3^1S_{0}$ & $0^{-+}$  & 10.372 && 10.342 &10.336 &10.367 &10.326& 10.353 &10.330& 10.314& 10.330&10.088&10.329&10.315\\  
$3^3S_{1}$  & $1^{--}$ & 10.375 & $10.3551\pm0.0005$ ($\Upsilon(3S)$) &10.354&10.354 &10.392 &10.343& 10.373&10.349& 10.334& 10.351&10.097&10.355&10.352\\
\noalign{\smallskip}  
$4^1S_{0}$ & $0^{-+}$ & 10.670 && 10.638 &10.523 &10.692 &10.584&&&& 10.595&&10.573&10.583\\  
$4^3S_{1}$  & $1^{--}$& 10.672 & $10.5794\pm0.0012$ ($\Upsilon(4S)$) & 10.650 &10.635 &10.705 &10.597& 10.654 &10.607& & 10.611&&10.586&10.615\\  
\noalign{\smallskip}
$5^1S_{0}$ & $0^{-+}$  & 10.928 && 10.901 &10.869 &10.981 &10.800& && & 10.817&&10.851&10.816\\   
$5^3S_{1}$  & $1^{--}$ & 10.930 & $10.8852^{+0.0026}_{-0.0016}$ ($\Upsilon(10860)$) & 10.912 &10.878 &10.991 &10.811& 10.999 &10.818&& 10.831&&10.869&10.845\\ 
\noalign{\smallskip} 
$6^1S_{0}$ &$0^{-+}$&11.158&&&11.097&&10.997&&&&&&11.061&11.024\\
$6^3S_{1}$&$1^{--}$&11.160& $11.000\pm0.004$ ($\Upsilon(11020)$) &&11.102&&10.988&&10.995&&&&11.088&11.051\\
 \noalign{\smallskip}\hline\noalign{\smallskip}
$1^3P_{0}$ &$0^{++}$&9.877 & $9.85944\pm0.00042\pm0.00031$ ($\chi_{b0}(1P)$) & 9.874 &9.847 &9.857 &9.804& 9.851 &9.855& 9.865& 9.865&9.755&9.859&9.837\\
$1^3P_{1}$ &$1^{++}$&9.889 & $9.89278\pm0.00026\pm0.00031$ ($\chi_{b1}(1P)$) & 9.894 &9.876 &9.893 &9.903& 9.890 &9.874& 9.893& 9.897&9.768&9.892&9.852\\
$1^1P_{1}$ & $1^{+-}$ & 9.902 & $9.8993\pm0.0008$ ($h_b(1P)$) & 9.899 &9.882 &9.899 &9.909& 9.886 &9.879& 9.900& 9.903&9.775&9.900&9.864\\
$1^3P_{2}$ & $2^{++}$ &9.908& $9.91221\pm0.00026\pm0.00031$ ($\chi_{b2}(1P)$) & 9.907 &9.897 &9.919 &9.921& 9.899 &9.886& 9.913& 9.918&9.792&9.912&9.877\\
\noalign{\smallskip}
$2^3P_{0}$ & $0^{++}$ & 10.268 & $10.2325\pm0.0004\pm0.0005$ ($\chi_{b0}(2P)$) &10.248 &10.226 &10.244 &10.226& 10.233 &10.221& 10.194& 10.226&10.035&10.233&10.258\\
$2^3P_{1}$ & $1^{++}$& 10.277 & $10.25546\pm0.00022\pm0.00050$ ($\chi_{b1}(2P)$) &10.265 &10.246 &10.276 &10.249& 10.257 &10.236& 10.212& 10.251&&10.255&10.279\\
$2^1P_{1}$ &$1^{+-}$ &10.286 & $10.2598\pm0.0005\pm0.0011$ ($h_b(2P)$) &10.268 &10.250 &10.269 &10.254& 10.262 &10.240& 10.219& 10.256&10.053&10.260&10.298\\
$2^3P_{2}$ & $2^{++}$ &  10.291 & $10.26865\pm0.00022\pm0.00050$ ($\chi_{b2}(2P)$) &10.274 &10.261 &10.287 &10.264& 10.274 &10.246& 10.227& 10.269&10.262&10.268&10.317\\
\noalign{\smallskip}
$3^3P_{0}$ &$0^{++}$& 10.577 &&10.551 &10.522 &10.566 &10.490& 10.533 &10.500& & 10.502&&10.251&10.503\\
$3^3P_{1}$ &$1^{++}$& 10.585 & $10.5134\pm0.0007$ ($\chi_{b1}(3P)$) &10.567 &10.538 &10.590 &10.515& 10.556 &10.513& & 10.524&&10.541&10.529\\
$3^1P_{1}$ & $1^{+-}$ &10.593&  &10.570&10.541 &10.561 &10.519 &10.560& 10.516 && 10.529&&10.544&10.555\\
$3^3P_{2}$ & $2^{++}$ &10.597& $10.5240\pm0.0008$ ($\chi_{b2}(3P)$) &10.576&10.550 &10.600 &10.528 &10.568& 10.521 && 10.540&&10.550&10.580\\
\noalign{\smallskip}
$4^3P_{0}$ & $0^{++}$ & 10.842&&&10.775&&&&&&&&10.781&10.727 \\
$4^3P_{1}$ & $1^{++}$& 10.850&&&10.788&&&&&&&&10.802&10.756\\
$4^1P_{1}$ &$1^{+-}$ &10.857 &&&10.790&&&&&&&&10.804&10.785\\
$4^3P_{2}$ & $2^{++}$ & 10.861&&&10.798&&&&&&&&10.812&10.814\\
\noalign{\smallskip}
$5^3P_{0}$ & $0^{++}$ & 11.079&&&11.044&&&&&&&&&10.930 \\
$5^3P_{1}$ & $1^{++}$&  11.086 &&&11.014&&&&&&&&&10.962\\
$5^1P_{1}$ &$1^{+-}$  &11.093 &&&11.022&&&&&&&&&10.994\\
$5^3P_{2}$ & $2^{++}$  &11.096 &&&11.016&&&&&&&&&11.026\\
 \noalign{\smallskip}\hline\noalign{\smallskip}
$1^3D_{1}$ &$1^{--}$& 10.160 &&10.145 &10.138 &10.151 &10.146& 10.136 &10.117& 10.150& 10.145&&10.154&10.086\\
$1^3D_{2}$ &$2^{--}$&  10.163 & $10.1637\pm0.0014$ ($\Upsilon_2(1D)$) &10.149 &10.147 &10.163 &10.153& 10.141 &10.122& 10.161& 10.151&&10.161&10.123\\
$1^1D_{2}$ & $2^{-+}$ &10.162 &&10.149 &10.148 &10.161 &10.153&  &10.123& 10.163& 10.152&&10.163&10.140\\
$1^3D_{3}$ & $3^{--}$ &10.164&&10.150 &10.155 &10.171 &10.157&  &10.127& 10.172& 10.156&&10.166&10.175\\
\noalign{\smallskip}
$2^3D_{1}$ & $1^{--}$ & 10.486 &&10.462 &10.441 &10.468 &10.425&10.454  &10.414& 10.458& 10.432&&10.435&10.451\\
$2^3D_{2}$ & $2^{--}$& 10.488&&10.465 &10.449 &10.479 &10.432&  &10.418& 10.401& 10.438&&10.443&10.497\\
$2^1D_{2}$ &$2^{-+}$ & 10.489 &&10.465 &10.450 &10.479 &10.432&&10.419  &10.447& 10.439&&10.445&10.519 \\
$2^3D_{3}$ & $3^{--}$ &10.488 &&10.466 &10.455 &10.486 &10.436&  &10.422& 10.459& 10.442&&10.449&10.563\\
\noalign{\smallskip}
$3^3D_{1}$ &$1^{--}$& 10.761 &&10.736 &10.698 &10.757 && 10.725 &10.653& & 10.670&&10.704&10.652\\
$3^3D_{2}$ &$2^{--}$&  10.763 &&10.740 &10.705 &10.766&& &10.657& & 10.676&&10.711&10.707\\
$3^1D_{2}$ & $2^{-+}$ &10.764 &&10.740 &10.706 &10.766 && & 10.658& &10.677&&10.713&10.733\\
$3^3D_{3}$ & $3^{--}$ & 10.764&&10.741 &10.711 &10.773 & && 10.660& &10.680&&10.717&10.787\\
\noalign{\smallskip}
$4^3D_{1}$ & $1^{--}$ & 11.005&&&10.928&&&&&&&&10.949&10.848 \\
$4^3D_{2}$ & $2^{--}$& 11.007&&&10.934&&&&&&&&10.957&10.909\\
$4^1D_{2}$ &$2^{-+}$ & 11.007&&&10.935&&&&&&&&10.963&10.980 \\
$4^3D_{3}$ & $3^{--}$ & 11.008&&&10.939&&&&&&&&10.959&11.000 \\
\noalign{\smallskip}
$5^3D_{1}$ &$1^{--}$& 11.225&&&&&&&&&&&&11.047 \\
$5^3D_{2}$ &$2^{--}$& 11.227&&&&&&&&&&&&11.113 \\
$5^1D_{2}$ & $2^{-+}$ & 11.228&&&&&&&&&&&&11.145 \\
$5^3D_{3}$ & $3^{--}$ & 11.227&&&&&&&&&&&&11.211 \\
 \noalign{\smallskip}\hline\noalign{\smallskip}
$1^3F_{2}$ &$2^{++}$&10.370 &&&10.350&&10.338&&10.315&&&&10.343&10.294\\
$1^3F_{3}$ &$3^{++}$&10.372&&&10.355&&10.340&&10.321&&&&10.346&10.355\\
$1^1F_{3}$ & $3^{+-}$ &10.372&&&10.358&&10.339&&10.322&&&&10.374&10.372\\
$1^3F_{4}$ & $4^{++}$ & 10.375&&&10.355&&10.340&&&&&&10.349&10.429\\
\noalign{\smallskip}
$2^3F_{2}$ & $2^{++}$& 10.661&&&10.615&&&&&&&&10.610&10.610\\
$2^3F_{3}$ & $3^{++}$& 10.662&&&10.619&&&&&&&&10.614&10.692\\
$2^1F_{3}$ &$3^{-+}$&10.662&&&10.619&&&&&&&&10.617&10.718 \\
$2^3F_{4}$ & $4^{++}$ & 10.665&&&10.622&&&&&&&&10.617&10.798\\
\noalign{\smallskip}
$3^3F_{2}$ &$2^{++}$&10.914 &\\
$3^3F_{3}$ &$3^{++}$&10.915&\\
$3^1F_{3}$ & $3^{+-}$ &10.915&&&\\
$3^3F_{4}$ & $4^{++}$ & 10.918&&\\
\noalign{\smallskip}
$4^3F_{2}$ & $2^{++}$& 11.142&&\\
$4^3F_{3}$ & $3^{++}$&  11.143&&&\\
$4^1F_{3}$ &$3^{-+}$& 11.143&& \\
$4^3F_{4}$ & $4^{++}$ &  11.145&&\\
\noalign{\smallskip}
$5^3F_{2}$ & $2^{++}$&  11.350&&\\
$5^3F_{3}$ & $3^{++}$& 11.351&&\\
$5^1F_{3}$ &$3^{-+}$& 11.351&&\\
$5^3F_{4}$ & $4^{++}$ &  11.353&&\\
 \noalign{\smallskip}\hline\noalign{\smallskip}
\end{tabular}

}
\end{center}
\end{table*}

\noindent The higher charmonium spectrum should be interpreted with caution.  The calculated $D$-wave states are in the region where $\psi(3770)$, $\psi_2(3823)$ and $\psi_3(3842)$ have been observed, which supports the qualitative placement of the first $D$ multiplet.  For the $2P$ and higher sectors, however, states such as $\chi_{c1}(3872)$ and structures near $3.9$--$4.3~\GeV$ may contain threshold or molecular components in addition to conventional quarkonium admixtures \cite{ExpX3872Belle,Aaij2013X3872,Uehara2006chic2,Brambilla2020XYZ}.  The upper charmonium entries in Table~\ref{Table:masses1} are therefore best viewed as the conventional $\ccbar$ reference spectrum against which such nontrivial structures can be judged.

\noindent In bottomonium the same spin-dependent interaction produces smaller fractional splittings.  The $\eta_b$--$\Upsilon$ separation and the $\chi_b$ fine structure are naturally reduced by the larger heavy-quark mass, and the calculated $1P$ and $2P$ levels remain close to the established states.  The experimental observations of the $\chi_b(3P)$ system, $h_b(1P,2P)$ and $\eta_b(1S,2S)$ provide particularly useful checks on this pattern \cite{Aad2012Chib,Aaij2014ChibProduction,ExpHBelle,ExpEtaBBaBar,Mizuk2012EtaB2S}.  The highest bottomonium entries should again be treated as conventional reference levels, especially because open-bottom thresholds and the observed $Z_b$ structures can modify the phenomenology near the upper part of the spectrum \cite{Garmash2016Zb,Brambilla2011QWG}.

\subsection{Regge trajectories}
\noindent The Regge analysis reorganizes the same mass tables into approximately linear relations in the $(M^2,J)$ and $(M^2,n_r)$ planes.  It is not an independent fit to new spectral input; rather, it is a compact diagnostic of whether the calculated states follow a smooth orbital and radial ordering.  The use of $M^2$ on the horizontal axis follows the Chew--Frautschi construction, while the separation into natural-parity, unnatural-parity, spin-resolved radial and spin-averaged radial panels makes it possible to distinguish central-potential trends from fine-structure effects \cite{ChewFrautschi1962,TangNorbury2000,Anisovich2000Regge,Li2004Regge,Sonnenschein2014Strings}.

\begin{figure}[H]
\centering
\captionsetup[subfigure]{font=footnotesize,labelfont=bf,skip=1pt}
\setlength{\abovecaptionskip}{3pt}
\setlength{\belowcaptionskip}{0pt}
\begin{subfigure}[t]{0.485\textwidth}
\centering
\includegraphics[width=\linewidth]{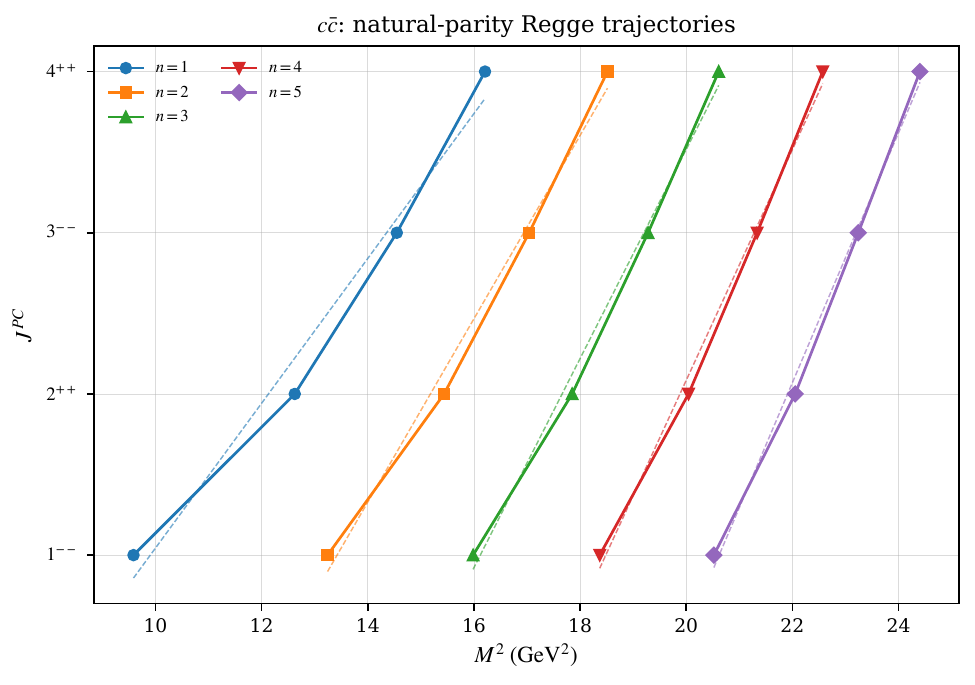}
\caption{Natural-parity $J^{PC}$ trajectory.}
\label{fig:srs-cc1}
\end{subfigure}\hfill
\begin{subfigure}[t]{0.485\textwidth}
\centering
\includegraphics[width=\linewidth]{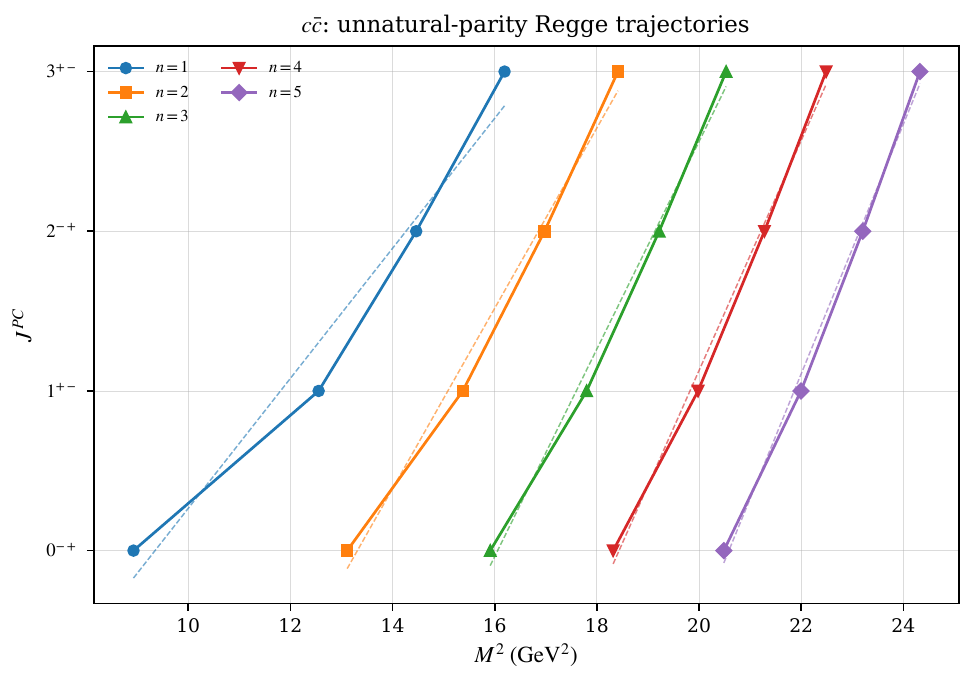}
\caption{Unnatural-parity $J^{PC}$ trajectory.}
\label{fig:srs-cc2}
\end{subfigure}

\vspace{0.30em}
\begin{subfigure}[t]{0.485\textwidth}
\centering
\includegraphics[width=\linewidth]{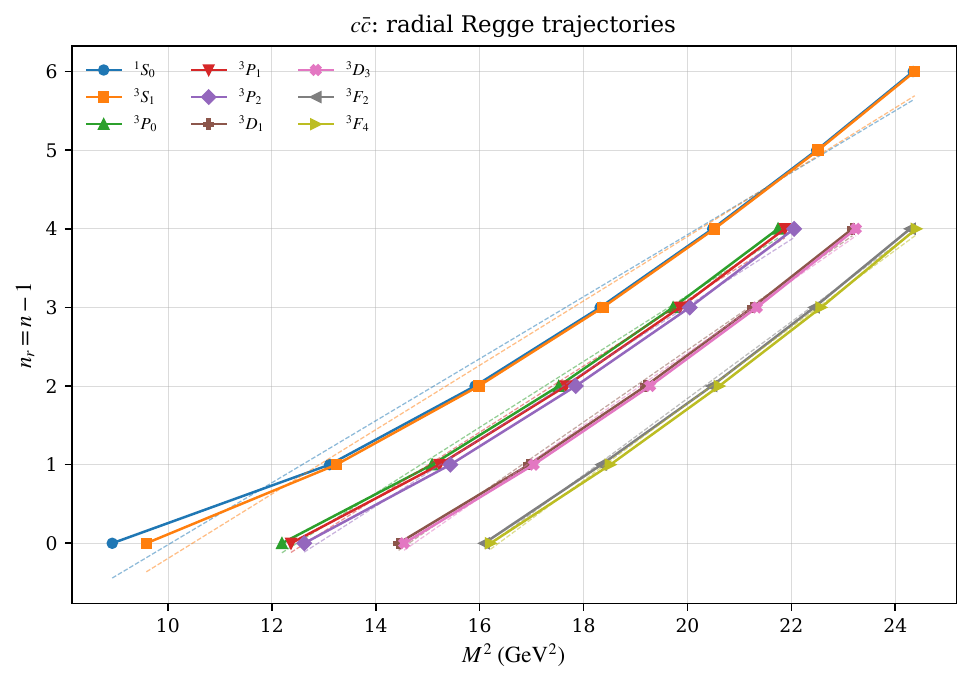}
\caption{Spin-resolved radial trajectories.}
\label{fig:srs-cc3}
\end{subfigure}\hfill
\begin{subfigure}[t]{0.485\textwidth}
\centering
\includegraphics[width=\linewidth]{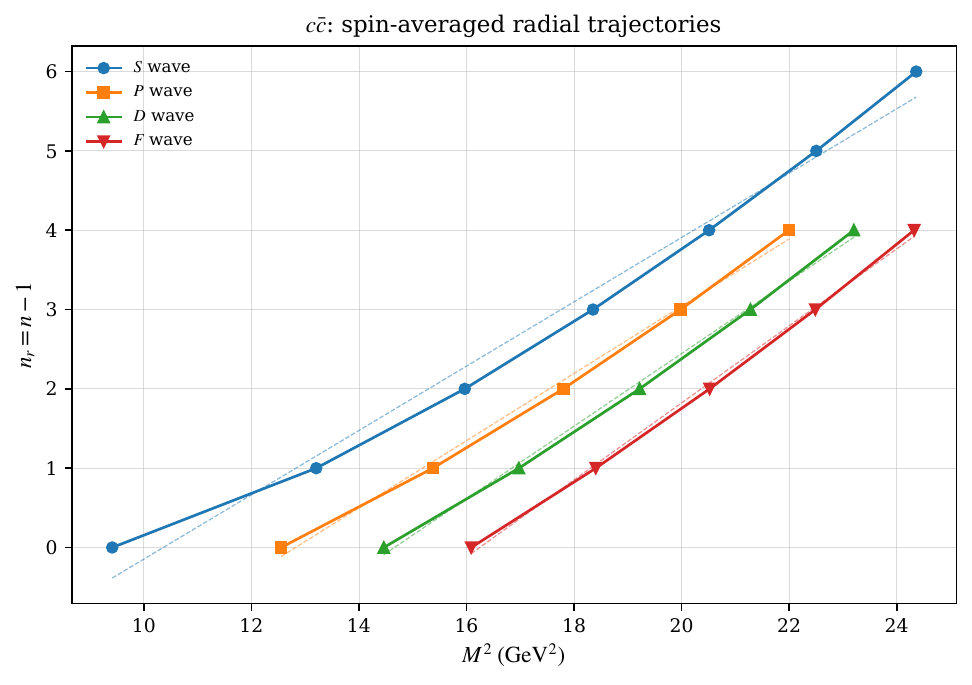}
\caption{Spin-averaged radial trajectories.}
\label{fig:srs-cc4}
\end{subfigure}
\caption{Regge trajectories for the $c\bar c$ meson.  The orbital panels use $M^2$ on the horizontal axis and $J^{PC}$ on the vertical axis, while the radial panels use $M^2$ on the horizontal axis and $n_r$ on the vertical axis.}
\label{fig:srs-cc-regge-panel}
\end{figure}

\begin{figure}[H]
\centering
\captionsetup[subfigure]{font=footnotesize,labelfont=bf,skip=1pt}
\setlength{\abovecaptionskip}{3pt}
\setlength{\belowcaptionskip}{0pt}
\begin{subfigure}[t]{0.485\textwidth}
\centering
\includegraphics[width=\linewidth]{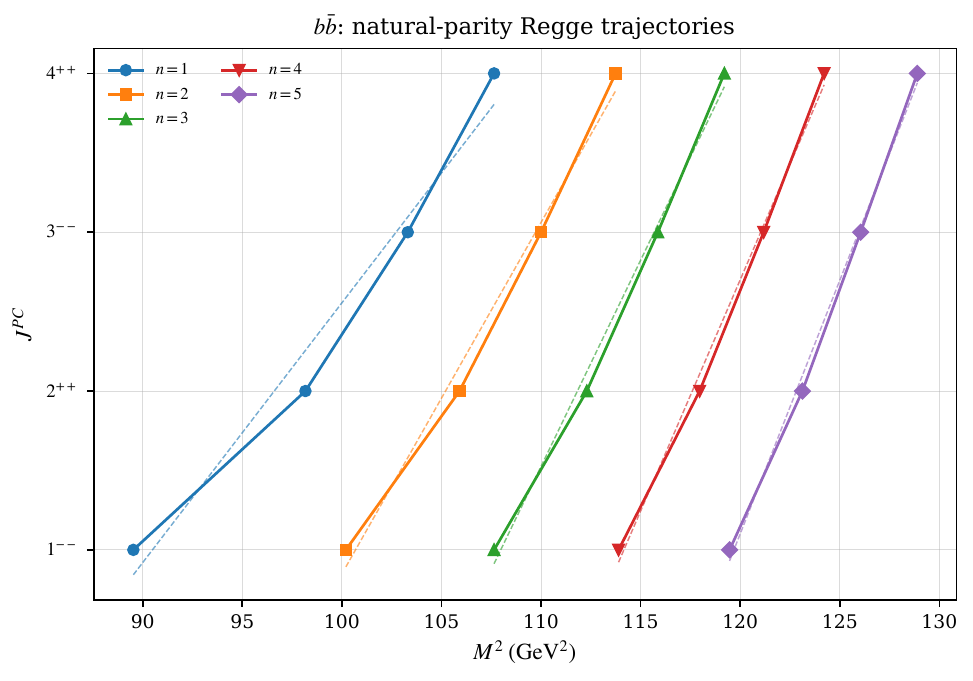}
\caption{Natural-parity $J^{PC}$ trajectory.}
\label{fig:srs-bb1}
\end{subfigure}\hfill
\begin{subfigure}[t]{0.485\textwidth}
\centering
\includegraphics[width=\linewidth]{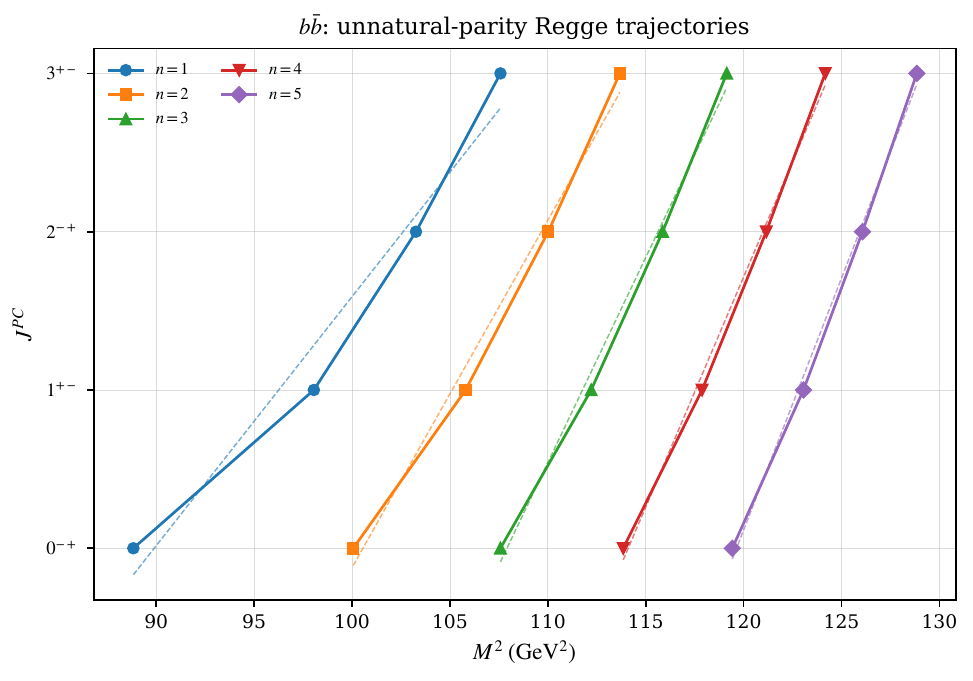}
\caption{Unnatural-parity $J^{PC}$ trajectory.}
\label{fig:srs-bb2}
\end{subfigure}

\vspace{0.30em}
\begin{subfigure}[t]{0.485\textwidth}
\centering
\includegraphics[width=\linewidth]{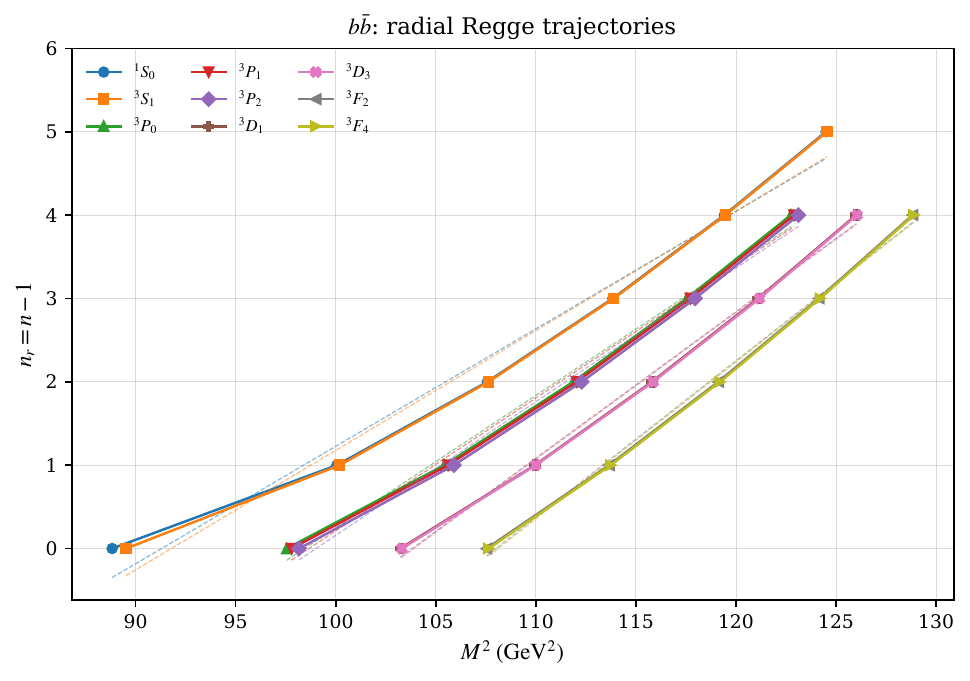}
\caption{Spin-resolved radial trajectories.}
\label{fig:srs-bb3}
\end{subfigure}\hfill
\begin{subfigure}[t]{0.485\textwidth}
\centering
\includegraphics[width=\linewidth]{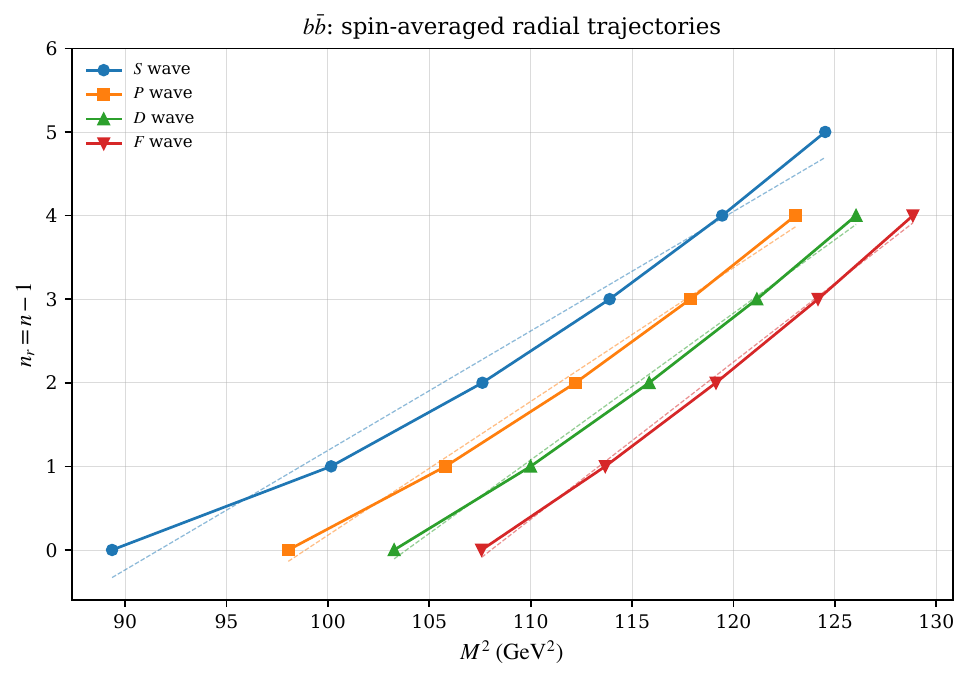}
\caption{Spin-averaged radial trajectories.}
\label{fig:srs-bb4}
\end{subfigure}
\caption{Regge trajectories for the $b\bar b$ meson.  The orbital panels use $M^2$ on the horizontal axis and $J^{PC}$ on the vertical axis, while the radial panels use $M^2$ on the horizontal axis and $n_r$ on the vertical axis.}
\label{fig:srs-bb-regge-panel}
\end{figure}

\begin{table*}[!htb]
\caption{Compact comparison of the $(M^{2},\,J)$ parent and daughter Regge-trajectory slopes and intercepts for $c\bar c$ and $b\bar b$ mesons.\label{Table:slope1}}
\centering
\resizebox{\textwidth}{!}{%
\begin{tabular}{llcccc}
\toprule
Parity & Trajectory & \multicolumn{2}{c}{$c\bar c$} & \multicolumn{2}{c}{$b\bar b$} \\
\cmidrule(lr){3-4}\cmidrule(lr){5-6}
 & & $\alpha(\mathrm{GeV}^{-2})$ & $\alpha_0$ & $\alpha(\mathrm{GeV}^{-2})$ & $\alpha_0$ \\
\midrule
Natural & Parent & $0.449\pm0.049$ & $-3.447\pm0.627$ & $0.171\pm0.026$ & $-14.53\pm2.663$ \\
Natural & $1^{st}$ daughter & $0.568\pm0.038$ & $-6.631\pm0.624$ & $0.310\pm0.093$ & $-30.379\pm9.930$ \\
Natural & $2^{nd}$ daughter & $0.648\pm0.037$ & $-9.456\pm0.690$ & $0.334\pm0.028$ & $-34.437\pm3.150$ \\
Natural & $3^{rd}$ daughter & $0.708\pm0.021$ & $-12.05\pm0.442$ & $0.336\pm0.017$ & $-35.441\pm2.012$ \\
Natural & $4^{th}$ daughter & $0.765\pm0.032$ & $-14.773\pm0.738$ & $0.441\pm0.052$ & $-49.612\pm6.140$ \\
\midrule
Unnatural & Parent & $0.411\pm0.050$ & $-3.834\pm0.970$ & $0.164\pm0.026$ & $-14.752\pm2.581$ \\
Unnatural & $1^{st}$ daughter & $0.558\pm0.030$ & $-7.408\pm0.486$ & $0.280\pm0.043$ & $-28.248\pm4.612$ \\
Unnatural & $2^{nd}$ daughter & $0.647\pm0.029$ & $-10.366\pm0.533$ & $0.342\pm0.032$ & $-36.255\pm3.573$ \\
Unnatural & $3^{rd}$ daughter & $0.716\pm0.026$ & $-13.173\pm0.547$ & $0.382\pm0.061$ & $-42.361\pm7.055$ \\
Unnatural & $4^{th}$ daughter & $0.783\pm0.029$ & $-16.105\pm0.672$ & $0.441\pm0.050$ & $-50.576\pm5.900$ \\
\bottomrule
\end{tabular}}
\end{table*}

\begin{table*}[!htb]
\caption{Compact comparison of the $(M^{2},\,n_r)$ Regge-trajectory slopes and intercepts for $c\bar c$ and $b\bar b$ mesons.\label{Table:slope2}}
\centering
\resizebox{\textwidth}{!}{%
\begin{tabular}{llcccccc}
\toprule
State & $J^{PC}$ & \multicolumn{3}{c}{$c\bar c$} & \multicolumn{3}{c}{$b\bar b$} \\
\cmidrule(lr){3-5}\cmidrule(lr){6-8}
 & & Meson & $\beta(\mathrm{GeV}^{-2})$ & $\beta_0$ & Meson & $\beta(\mathrm{GeV}^{-2})$ & $\beta_0$ \\
\midrule
S & $0^{-+}$ & $\eta_c$ & $0.346\pm0.027$ & $-2.31\pm0.440$ & $\eta_b$ & $0.155\pm0.021$ & $-14.160\pm2.189$ \\
S & $1^{--}$ & $J/\psi$ & $0.366\pm0.022$ & $-2.687\pm0.367$ & $\Upsilon$ & $0.160\pm0.023$ & $-14.173\pm2.294$ \\
P & $0^{++}$ & $\chi_{c0}$ & $0.419\pm0.017$ & $-4.231\pm0.363$ & $\chi_{b0}$ & $0.205\pm0.019$ & $-20.351\pm2.099$ \\
P & $1^{++}$ & $\chi_{c1}$ & $0.427\pm0.017$ & $-4.513\pm0.289$ & $\chi_{b1}$ & $0.211\pm0.020$ & $-21.034\pm2.222$ \\
D & $1^{+-}$ & $h_c$ & $0.458\pm0.014$ & $-5.703\pm0.270$ & $h_b$ & $0.248\pm0.010$ & $-25.695\pm1.212$ \\
D & $2^{++}$ & $\chi_{c2}$ & $0.459\pm0.014$ & $-5.778\pm0.273$ & $\chi_{b2}$ & $0.247\pm0.011$ & $-25.617\pm1.274$ \\
F & $2^{++}$ & $c\bar c$ & $0.487\pm0.011$ & $-6.903\pm0.234$ & $b\bar b$ & $0.265\pm0.022$ & $-28.449\pm2.572$ \\
F & $4^{++}$ & $c\bar c$ & $0.485\pm0.011$ & $-6.939\pm0.233$ & $b\bar b$ & $0.241\pm0.025$ & $-25.552\pm2.959$ \\
\bottomrule
\end{tabular}}
\end{table*}

\begin{table*}[!htb]
\caption{Side-by-side comparison of the slopes and intercepts of the $(M^{2},\,n_r)$ center-weighted S-, P-, D- and F-state Regge trajectories for $c\bar{c}$ and $b\bar{b}$ mesons.\label{Table:slope3}}
\centering
\resizebox{0.82\textwidth}{!}{%
\begin{tabular}{lcccc}
\toprule
Trajectory & \multicolumn{2}{c}{$c\bar c$} & \multicolumn{2}{c}{$b\bar b$} \\
\cmidrule(lr){2-3}\cmidrule(lr){4-5}
 & $\beta(\mathrm{GeV}^{-2})$ & $\beta_{0}$ & $\beta(\mathrm{GeV}^{-2})$ & $\beta_{0}$ \\
\midrule
S State & $0.360\pm0.024$ & $-2.588\pm0.388$ & $0.159\pm0.020$ & $-13.562\pm2.161$ \\
P State & $0.423\pm0.016$ & $-4.432\pm0.294$ & $0.208\pm0.018$ & $-20.716\pm2.011$ \\
D State & $0.457\pm0.014$ & $-5.709\pm0.270$ & $0.247\pm0.010$ & $-25.647\pm1.209$ \\
F State & $0.485\pm0.012$ & $-6.886\pm0.263$ & $0.254\pm0.022$ & $-27.034\pm2.574$ \\
\bottomrule
\end{tabular}}
\end{table*}

\noindent Tables~\ref{Table:slope1} and \ref{Table:slope2} list the fitted parent and daughter slopes for the orbital and radial Regge families, respectively.  The charmonium slopes are larger than the bottomonium slopes in both orbital and radial fits.  This is a direct consequence of the lower reduced mass and larger excitation spacings in the $\ccbar$ system.  Bottomonium, being more compact and closer to the nonrelativistic limit, produces flatter trajectories in $M^2$.  The spin-averaged radial slopes in Table~\ref{Table:slope3} are especially useful because they reduce the visual impact of fine splittings and expose the underlying central-potential trend.  The comparison with Regge studies of heavy-light mesons, $B_c$ mesons, baryons and multiquark systems is methodological rather than numerical: those works show that trajectory slopes are sensitive to constituent mass, screening and threshold effects, while the present hidden-flavour systems provide a cleaner two-body benchmark \cite{Oudichhya:2024hmn,Patel:2025rsf,Jakhad:2025zos,Patel:2026prg}.

\subsection{E1 transition}
\noindent The E1 widths in Tables~\ref{Table:ccE1} and \ref{Table:bbE1} are controlled by the cubic photon-energy factor, the angular coefficient and the radial dipole matrix element.  The strongest channels are generally those connecting nearby orbital multiplets with large radial overlap, such as $1P\to1S$ and $1D\to1P$.  Node-sensitive transitions involving radially excited states show a much larger model spread, because small differences in the wave-function phase and the relative placement of nodes can strongly alter the overlap integral.  For this reason the E1 widths are more discriminating than masses alone: two models can give similar spectra but noticeably different dipole widths.

\begin{table*}[!htb]
\begin{center}
\setlength{\tabcolsep}{2pt}
\renewcommand{\arraystretch}{1.06}
{\caption{\label{Table:ccE1}{E1 transition widths for $c\bar{c}$ mesons.}}}

\resizebox{\textwidth}{!}{
\begin{tabular}{c c c c c c c c c c c c c c c}
\noalign{\smallskip}\hline\noalign{\smallskip}
\noalign{\smallskip}\hline\noalign{\smallskip}
 State &&& Recent study && &&Another study ($\Gamma$ in KeV) \\ 
\noalign{\smallskip}\hline\noalign{\smallskip}
 Initial & Final & $ E_{\gamma}(MeV) $ & $\Gamma(KeV)$ &Expt\cite{pdg2}  & Ref.\cite{N. Devlani} &Ref.\cite{B.Q. Li1} &Ref.\cite{D. Ebert1}&Ref.\cite{A. Parmar}&Ref.\cite{N.Brambilla}&Ref.\cite{Raghav Chaturvedi1}&Ref.\cite{L. Cao}&Ref.\cite{W.J. Deng}&Ref.\cite{M.A. Sultan}&Ref.\cite{N . R. Soni} \\ 
\noalign{\smallskip}\hline\noalign{\smallskip}
 $(1^3P_{2})$& $(1^3S_{1})$&427&326.88&406$\pm$31&233.85 &309&327&383&315&457.39&405&338&424.5&157.22\\
  $(1^{3}P_{1})$&$(1^3S_{1})$&395&300.62&320$\pm$25&189.86 &244&265&361&241&378.33&341&278&319.5&146.32\\
  $(1^{1}P_{1})$&$(1^1S_{0})$&511&217.20&&357.83 &323&560&671&482&5050.69&473&373&490.3&247.97\\
   $(1^3P_{0})$&$(1^3S_{1})$&373&84.53&131$\pm$14&118.29 &117&121&264&120&145.33&104&146&154.5&112.03\\
 
  \noalign{\smallskip}\hline\noalign{\smallskip}
  $(2^3S_{1})$ & $(1^3P_{2})$&30.084&34.89&26$ \pm $1.5  &7.07&34&18.2&&30.1&28.51&39&44&37.9&62.31\\
  $(2^3S_{1})$&$(1^{3}P_{1})$&40.94&48.3&27.9$\pm$1.5  &10.39&36&22.9&&42.8&28.79                &38&48&54.2&43.29\\
  $(2^3S_{1})$&$(1^{1}P_{1})$&14.44&3.79&  &7.94&104& &&&28.9                    \\
  $(2^3S_{1})$&$(1^3P_{0})$&105&30.98&29.8$\pm$1.5  &11.93&25&26.3&&47&31&29&26&62.6&21.86                     \\
  $(2^1S_{0})$&$(1^{3}P_{1})$&19&9.03& &9.02&                     &56&52&49.9&36.20\\
  $(2^1S_{0})$&$(1^{1}P_{1})$&&&&6.05&&6.2&&35.1 &2.69                    \\
  \noalign{\smallskip}\hline\noalign{\smallskip}
  $(1^3D_{3})$&$(1^3P_{2})$&265&185.7&  &237.5&323&156&432&402&348.99&302&&271.1&171.21\\
  $(1^{3}D_{2})$&$(1^3P_{2})$&257&42.1&  &62.34&55&59&131&69.5&66.92&82&82&64.06&50.31\\
  $(1^{3}D_{2})$&$(1^3P_{1})$&330&26.75&  &86.32&208&215&423&313&103.70&301&291&311.2&165.17\\
  $(1^3D_{1})$&$(1^3P_{2})$&238&22&$<$21  &6.45&4.6&6.9&15.2&3.88&13.13&8.1&5.7&4.86&5.72\\
  $(1^3D_{1})$&$(1^{3}P_{1})$&272&63.20& 70$\pm$17 &139.52&93&135&246&99&90.47&153&111&126.2&93.77\\
  $(1^3D_{1})$&$(1^1P_{1})$&295&180.64&172 $\pm$30 &343.87&197&355&448&299&120.66&362&232&405.4&161.50\\
  $(1^3D_{1})$&$(1^3P_{0})$&268&79.7&  &124.14&&&&&\\
 \noalign{\smallskip}\hline\noalign{\smallskip}
  $(2^3P_{2})$&$(2^3S_{1})$&376&277.23&  &281.93&100&&164&&346.02&264&&287.5&116.32\\
  $(2^{3}P_{1})$&$(2^3S_{1})$&332&253.2&  &206.57&60&&174&&219.71&234&&185.3&102.67\\
  $(2^{1}P_{1})$&$(2^1S_{0})$&402&494.41&  &343.55&108&&333&&493.45&274&&272.9&163.64\\
  $(2^3P_{0})$&$(2^3S_{1})$&302&140.28&  &102.23&44&&112&&161.07&83&&108.3&70.40\\
   \noalign{\smallskip}\hline\noalign{\smallskip}
  $(2^3P_{2})$&$(1^3D_{3})$&103&22.9&  &33.27&&&&&6.88&76&&78.69&\\

  $(2^3P_{2})$&$(1^3D_{2})$&112&5.21&  &5.49&&&&&7.47&10&&15.34&\\

  $(2^3P_{2})$&$(1^1D_{2})$&112&5.21&  &5.83&&&&&9.63&0.064&&1.67&\\

  $(2^{3}(P_{2})$&$(1^3D_{1})$&126&0.73&  &0.41&&&&&9.07&11&&21.53&\\

  $(2^{3}P_{1})$&$(1^3D_{1})$&79&3.04&  &5.35&&&&&4.19&1.4&&13.55\\
  $(2^3P_{0})$&$(1^3D_{1})$&47&2.66&  &3.21&&&&&2.31\\
  \noalign{\smallskip}\hline\noalign{\smallskip}
\end{tabular}
}

\end{center}
\end{table*}

\begin{table*}[!htb]
\begin{center}
\setlength{\tabcolsep}{2pt}
\renewcommand{\arraystretch}{1.06}
{\caption{\label{Table:bbE1}{E1 transition widths for $b\bar{b}$ mesons.}}}

\resizebox{\textwidth}{!}{
\begin{tabular}{c c c c c c c c c c c c c c c}
\noalign{\smallskip}\hline\noalign{\smallskip}
\noalign{\smallskip}\hline\noalign{\smallskip}
 State && Recent study &&&&Another study ($\Gamma$ in KeV) \\ 
\noalign{\smallskip}\hline\noalign{\smallskip}
 Initial & Final & $ E_{\gamma}(MeV) $ & $\Gamma(KeV)$ &EXP\cite{pdg}  & Ref.\cite{N. Devlani}&Ref.\cite{D. Ebert1}&Ref.\cite{N. Akbar}&Ref.\cite{S. Godfrey1} &Ref.\cite{A. Parmar}&Ref.\cite{N.Brambilla}&Ref.\cite{Raghav Chaturvedi}&Ref. \cite{W.J. Deng}&Ref.  \cite{B.Q. Li}  \\  
\noalign{\smallskip}\hline\noalign{\smallskip}
 $(1^3P_{2})$&$(1^3S_{1})$&435&26.90&34.38&19.99  &40.2&37.3&32.8&70.29&31.6&39.266&31.8&38.2\\
  $(1^{3}P_{1})$&$(1^3S_{1})$&430&23.44 &32.54&18.40&36.6&32.8&29.5&67.43&27.8&42.551&31.9&33.6\\
  $(1^{1}P_{1})$&$(1^1S_{0})$&453&40.45&35.77  &24.68&52.64&22.9&35.7&94.05&41.8&74.633&35.8&35.8\\
   $(1^3P_{0})$&$(1^3S_{1})$&435&21.54&  &16.0&29.9&26.0&23.8&58.42&22.2&29.08&27.5&26.6\\
 
\noalign{\smallskip}\hline\noalign{\smallskip}
  $(2^3S_{1})$&$(1^3P_{2})$&101&1.82&2.29$ \pm $0.20  &0.58&2.46&2.92&1.88&&2.04&2.274&2.62&2.62\\
  $(2^3S_{1})$&$(1^{3}P_{1})$&107&2.6&2.21$\pm$0.19  &0.51&2.45&2.83&1.63&&2.00 &2.232&2.17&2.54\\                 
  $(2^3S_{1})$&$(1^{1}P_{1})$&120&1.24&  &&&&&&&2.274&2362&2.62\\
  $(2^3S_{1})$&$(1^3P_{0})$&132&1.50& 1.22$ \pm $0.11  &0.29&1.62&1.85&0.91&&1.29&1.300&1.09&1.67\\
  $(2^1S_{0})$&$(1^{3}P_{1})$&56&5.52&  &0.94&\\
  $(2^1S_{0})$&$(1^{1}P_{1})$&42&2.37&  &0.80&3.09&4.54&2.48&&41.8\\
\noalign{\smallskip}\hline\noalign{\smallskip}
  $(1^3D_{3})$&$(1^3P_{2})$&20.39&3.64&  &21.68&24.6&31.27&24.3&76.82&22.6&5.009&350&284\\
  $(1^{3}D_{2})$&$(1^3P_{2})$&20&8.72&  &5.34&6.35&5.48&5.6&20.60&5.46&1.908&1.02&0.65\\
  $(1^{3}D_{2})$&$(1^3P_{1})$&22.25&3.55&  &6.19&23.3&21.03&19.2&65.35&20.5&3.940&21.8&23.8\\
  $(1^3D_{1})$&$(1^3P_{2})$&19.19&0.94&  &0.57&0.69&0.68&0.56&2.32&0.50&1.908&1.02&0.65\\
  $(1^3D_{1})$&$(1^{3}P_{1})$&22.05&1.92&  &9.91&12.7&10.96&9.7&36.75&10.7\\
  $(1^3D_{1})$&$(1^1P_{1})$&20.68&1.58&  &9.3&49&52.2&\\
  $(1^3D_{1})$&$(1^3P_{0})$&23.32&3.03&&16.58  &23.4&25.84&16.5&5.847&20.1&3.120&19.8&23.6\\
 \noalign{\smallskip}\hline\noalign{\smallskip}
  $(2^3P_{2})$&$(2^3S_{1})$&277&28.9&24.645  &20.02&16.7&18.99&14.3&32.92&14.5&15.665&15.5&15.9\\
  $(2^{3}P_{1})$&$(2^3S_{1})$&465&24.3&23.283  &18.03&14.7&16.14&13.3&30.82&12.4&21.844&15.3&15.9\\
  $(2^{1}P_{1})$&$(2^1S_{0})$&270&48.31&40.32  &22.09&21.4&12.79&14.1&&19&23.106&16.24.7\\
  $(2^3P_{0})$&$(2^3S_{1})$&254&0.002&0.00012  &14.96&6.79&11.88&10.9&25.40&9.17&16.842&14.4&11.7\\
   \noalign{\smallskip}\hline\noalign{\smallskip}
  $(2^3P_{2})$&$(1^3D_{3})$&8.46&0.142&  &1.55&2.35&2.885&1.5&&2.25\\

  $(2^3P_{2})$&$(1^3D_{2})$&8.76&0.281&  &0.28&0.449&1.054&0.3&&0.434\\

  $(2^3P_{2})$&$(1^1D_{2})$&8.66&0.272&  &0.29\\

  $(2^{3}(P_{2})$&$(1^3D_{1})$&8.96&0.301&  &0.02&0.35&0.79&0.03&&0.34\\

  $(2^{3}P_{1})$&$(1^3D_{1})$&7.57&3.03&  &0.41&0.615&1.46&0.5&&0.593\\

  $(2^3P_{0})$&$(1^3D_{1})$&6.67&8.31&  &1.06&1.17&3.21&1.0&&1.13\\

  \noalign{\smallskip}\hline\noalign{\smallskip} 
\end{tabular}
}

\end{center}
\end{table*}
\clearpage

\noindent The charmonium E1 widths are typically larger because the photon energies and dipole overlaps are less suppressed.  The comparison with earlier radiative-transition calculations shows that the leading low-lying transitions are relatively stable across methods, whereas hindered or node-sensitive transitions vary substantially \cite{KwongRosner1988Dwave,Ackleh1996StrongDecay,Ganbold2021CharmoniumRadiative,Delaney2024LatticeRadiative}.  In bottomonium the same selection rules apply, but the heavier reduced mass and the smaller bottom-quark electric charge reduce many rates.  The $E_\gamma^3$ factor also makes transitions with small mass gaps very sensitive to the details of the spectrum.  The E1 tables therefore provide a direct consistency test between the calculated masses, photon energies and wave-function overlaps.

\subsection{M1 transition}
\noindent The M1 widths in Tables~\ref{Table:ccm1} and \ref{Table:bbm1} test a different part of the wave functions.  Allowed $n^3S_1\to n^1S_0$ transitions are mainly governed by the hyperfine splitting and the magnetic moment factor, while hindered transitions between different radial levels depend on cancellations in the spherical-Bessel overlap.  This makes M1 transitions particularly sensitive to relativistic corrections and to the detailed radial structure.  The $J/\psi\to\eta_c\gamma$ channel is the most useful benchmark because it has been studied experimentally, in lattice QCD and in light-cone sum rules \cite{Donald2012JpsiLattice,Delaney2024LatticeRadiative,Guo2020LCSRJpsiEtaC}.

\begin{table*}[!htb]
\begin{center}
\setlength{\tabcolsep}{2pt}
\renewcommand{\arraystretch}{1.06}
{\caption{ \label{Table:ccm1} {M1 transition widths for $c\bar{c}$ mesons.}}}

\resizebox{\textwidth}{!}{
\begin{tabular}{c c c c c c c c c c c c c c c}
\noalign{\smallskip}\hline\noalign{\smallskip}
\noalign{\smallskip}\hline\noalign{\smallskip}
 State&& Recent study &&& &Another study ($\Gamma$ in KeV)  \\ 
\noalign{\smallskip}\hline\noalign{\smallskip} 
 Initial $\rightarrow$ Final & $ E_{\gamma}(MeV) $ & $\Gamma(KeV)$&Expt\cite{pdg2}   & Ref.\cite{N. Devlani}&Ref.\cite{D. Ebert1}&Ref.\cite{A. Parmar}&Ref.\cite{N.Brambilla}&Ref.\cite{Raghav Chaturvedi1}&Ref.\cite{L. Cao}&Ref.\cite{W.J. Deng}&Ref.\cite{M.A. Sultan}&Ref.\cite{N . R. Soni}\\  
\noalign{\smallskip}\hline\noalign{\smallskip}
 $(1^3S_{1})$ $\rightarrow$ $(1^1S_{0})$&38&1.33&1.58$\pm$.037 &1.647&1.05&2.01&1.960&1.255&2.2&2.44&2.752&1.18\\
  $(2^3S_{1})$ $\rightarrow$ $(2^1S_{0})$&43&0.65&0.21$\pm$0.15 &0.135&0.99&0.20&0.140&1.350&0.096&0.19&0.197&0.50\\
   $(3^3S_{1})$ $\rightarrow$ $(3^1S_{0})$&159&0.457&&0.135&&0.120&&1.194&0.044&0.088&0.044&0.36\\
    $(2^3S_{1})$ $\rightarrow$ $(1^1S_{0})$&892&1.84&1.24$\pm$0.29&69.57&0.95&&0.926&0.846&3.8&7.80&4.532&3.25\\
     $(2^1S_{0})$ $\rightarrow$ $(1^3S_{0})$&462&0.88 &107.2&1.12&&0.538&&4.060&6.9&2.29&7.962\\

\noalign{\smallskip}\hline\noalign{\smallskip}
  $(1^3P_{2})$ $\rightarrow$ $(1^3P_{0})$&129&0.240&&1.638&&&&8.043\\
   $(1^3P_{2})$ $\rightarrow$ $(1^3P_{1})$&78&0.160&&0.189&&&&1.592\\
    $(1^3P_{2})$ $\rightarrow$ $(1^1P_{1})$&25&0.0586&&0.056&&&&0.245\\
     $(1^1P_{1})$ $\rightarrow$ $(1^3P_{0})$&10.44&0.126&&0.782&&&&2.768\\
\noalign{\smallskip}\hline\noalign{\smallskip}
 \end{tabular}
 }

\end{center}
\end{table*}

\begin{table*}[!htb]
\begin{center}
\setlength{\tabcolsep}{2pt}
\renewcommand{\arraystretch}{1.06}
{\caption{ \label{Table:bbm1} {M1 transition widths for $b\bar{b}$ mesons.}}}

\resizebox{\textwidth}{!}{
\begin{tabular}{c c c c c c c c c c c c c c}
\noalign{\smallskip}\hline\noalign{\smallskip}
\noalign{\smallskip}\hline\noalign{\smallskip}
 State& Recent study &&  &&Another study ($\Gamma$ in KeV) \\ 
\noalign{\smallskip}\hline\noalign{\smallskip}
 Initial $\rightarrow$ Final & $ E_{\gamma}(MeV) $ & $\Gamma(KeV)$ &EXP\cite{pdg}  & Ref.\cite{N. Devlani}&Ref.\cite{D. Ebert1}&Ref.\cite{N. Akbar}&Ref.\cite{S. Godfrey1} &Ref.\cite{A. Parmar}&Ref.\cite{N.Brambilla}&Ref.\cite{Raghav Chaturvedi}&Ref. \cite{W.J. Deng}&Ref.  \cite{S. F. Radford}\\  
\noalign{\smallskip}\hline\noalign{\smallskip}
 $(1^3S_{1})$ $\rightarrow$ $(1^1S_{0})$&36&0.005&&0.003&0.011&0.010&5.8&0.0182&0.00895&2.527&10&4.0\\
  $(2^3S_{1})$ $\rightarrow$ $(2^1S_{0})$&69&0.0000388&&0.0002&0.00066&0.00059&1.4&0.00401&0.00151&
  0.306&0.59&0.05\\
   $(3^3S_{1})$ $\rightarrow$ $(3^1S_{0})$&29&0.00003&&0.00005&0.012&0.00025&0.8&0.00086&0.00083&0.024&3.9&\\
    $(2^3S_{1})$ $\rightarrow$ $(1^1S_{0})$&583&6.4&12.5$\pm$4.9&1.61&0.1729&0.081&6.4&&0.00281&11.954&66&0.0\\
     $(2^1S_{0})$ $\rightarrow$ $(1^3S_{0})$&501&1.159&&3.59&0.00064&0.068&11.8&&0.00283&\\
\noalign{\smallskip}\hline\noalign{\smallskip}
 
 \end{tabular}
 }

\end{center}
\end{table*}
\clearpage

\noindent The calculated hierarchy follows the expected heavy-quark pattern.  Charmonium M1 widths are larger because the charm magnetic moment and hyperfine gaps are larger.  Bottomonium allowed M1 widths are strongly suppressed, and some hindered bottomonium channels become comparable or larger only when the photon energy is sufficiently large.  This behavior is consistent with NRQCD and pNRQCD treatments, where current matching, relativistic corrections and color-octet effects can be especially important for hindered M1 transitions \cite{Bodwin1995NRQCD,Brambilla2006M1,Jia2010HinderedM1}.

\subsection{Finite-spectrum thermodynamic observables}
\begin{figure}[H]
\centering
\includegraphics[width=0.92\textwidth]{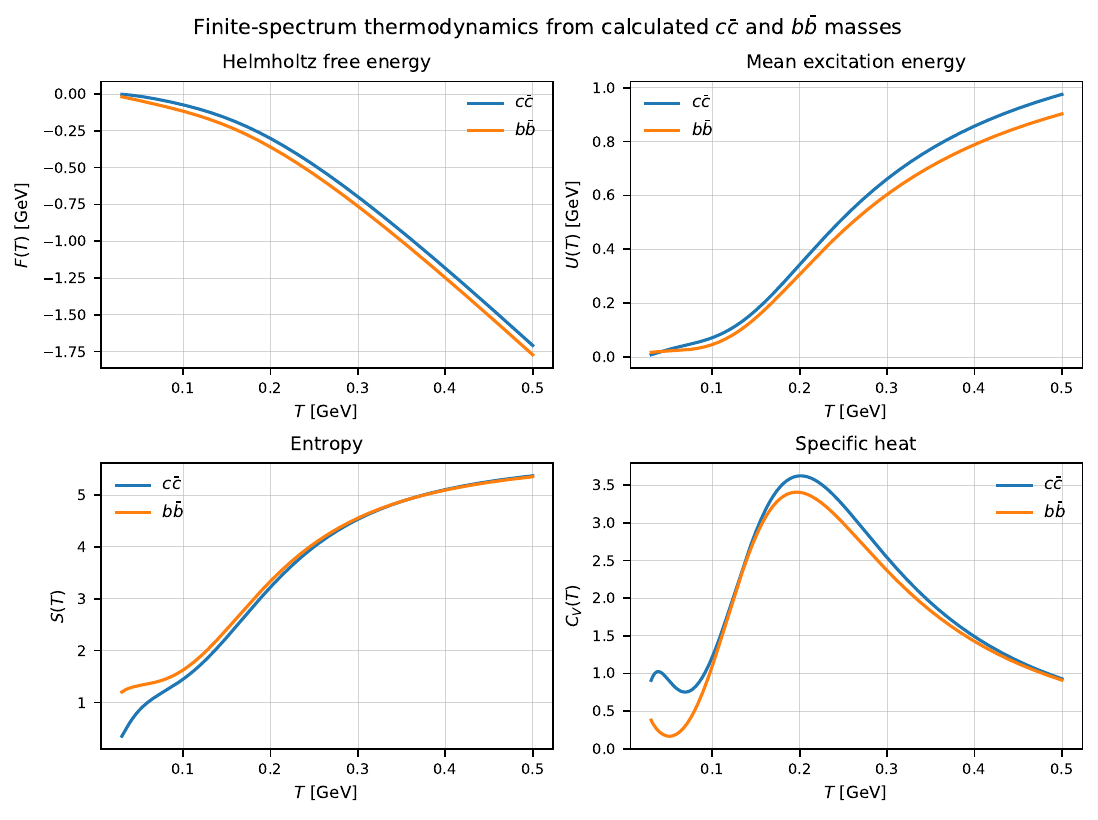}
\caption{Finite-spectrum thermodynamic observables derived from the calculated hidden-charm and hidden-bottom mass spectra through $\epsilon_i=M_i-M_0$.  The use of excitation energies isolates the effect of level spacing and spin degeneracy; it does not include explicit in-medium mass shifts, magnetic-field response, or continuum spectral modifications.}
\label{fig:finite_spectrum_thermo}
\end{figure}

\noindent The finite-spectrum thermodynamic analysis uses the calculated spin-resolved masses as a discrete set of levels.  The excitation energies are measured from the corresponding calculated ground states, $M_0=2.988~\GeV$ for charmonium and $M_0=9.425~\GeV$ for bottomonium.  The sums include 74 $\ccbar$ states over $2.988$--$4.940~\GeV$ and 72 $\bbbar$ states over $9.425$--$11.353~\GeV$, with each level weighted by $2J+1$.  This convention removes the trivial heavy-quark rest-mass offset and allows the plotted observables to reflect level spacing and spin multiplicity.  The representative numerical values are listed in Table~\ref{tab:finite_thermo}, and the full temperature dependence is shown in Fig.~\ref{fig:finite_spectrum_thermo}.

\begin{table}[H]
\caption{Finite-spectrum thermodynamic observables obtained from the calculated $c\bar c$ and $b\bar b$ spin-resolved mass spectra.  The free energy $F$ and mean excitation energy $U$ are in GeV, while $S$ and $C_V$ are dimensionless in units with $k_B=1$.}
\label{tab:finite_thermo}
\centering
\resizebox{\textwidth}{!}{%
\begin{tabular}{c|rrrr|rrrr}
\toprule
$T$ & \multicolumn{4}{c|}{$c\bar c$} & \multicolumn{4}{c}{$b\bar b$} \\
\cmidrule(lr){2-5}\cmidrule(lr){6-9}
(GeV) & $F$ & $U$ & $S$ & $C_V$ & $F$ & $U$ & $S$ & $C_V$ \\
\midrule
0.10 & -0.073 & 0.071 & 1.449 & 1.218 & -0.117 & 0.046 & 1.626 & 1.099\\
0.15 & -0.165 & 0.174 & 2.256 & 2.887 & -0.216 & 0.146 & 2.412 & 2.824\\
0.20 & -0.302 & 0.343 & 3.223 & 3.624 & -0.360 & 0.308 & 3.337 & 3.405\\
0.30 & -0.698 & 0.661 & 4.530 & 2.537 & -0.762 & 0.604 & 4.553 & 2.365\\
0.40 & -1.183 & 0.857 & 5.101 & 1.488 & -1.248 & 0.789 & 5.092 & 1.426\\
\bottomrule
\end{tabular}}
\end{table}

\noindent The free energy becomes negative because the ground-state excitation energy is set to zero and the degeneracy-weighted partition function exceeds unity once excited states contribute.  The low-temperature bottomonium curve is affected early by the nearby $\Upsilon(1S)$ spin partner, while charmonium rises more gradually because the first excitation gap is larger.  At higher temperature, additional radial and orbital states enter both sums and increase the entropy and specific heat.  These quantities should not be interpreted as a full hot-QCD equation of state, a deconfinement calculation or a magnetic-field response.  They are finite-spectrum indicators, conceptually related to spectrum-based thermodynamic analyses \cite{AbuShadyFathAllah2025}, and distinct from hot-medium magnetic-susceptibility studies that require a magnetic-field-dependent pressure and active quark degrees of freedom \cite{SamantaBroniowski2026}.

\subsection{QCD sum-rule decay constants and annihilation widths}
\begin{figure}[H]
\centering
\begin{subfigure}{0.48\textwidth}
    \centering
    \includegraphics[width=\linewidth]{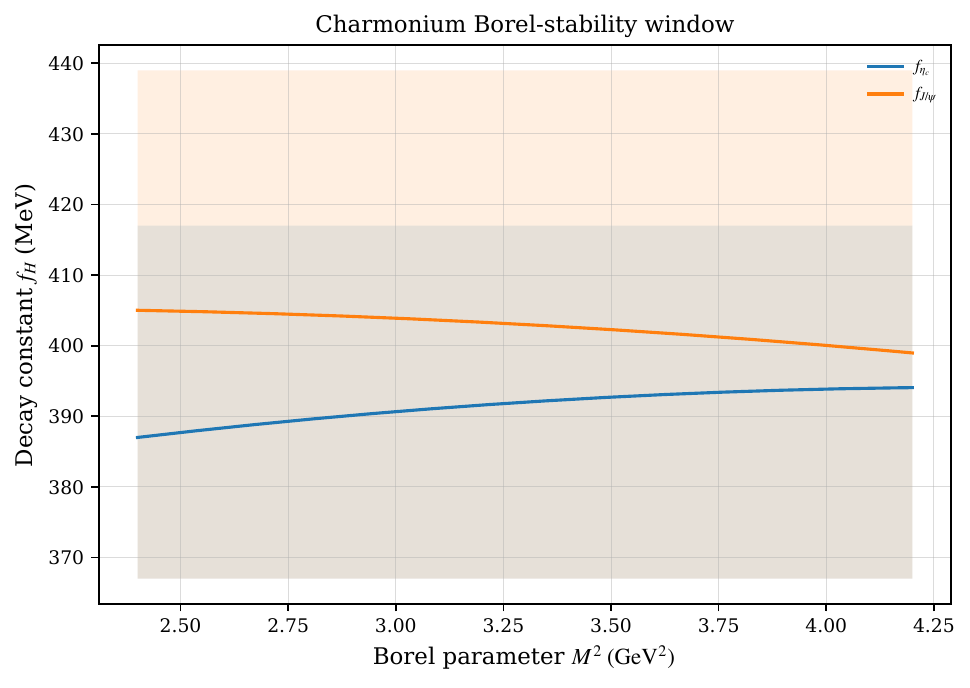}
    \caption{Charmonium.}
\end{subfigure}\hfill
\begin{subfigure}{0.48\textwidth}
    \centering
    \includegraphics[width=\linewidth]{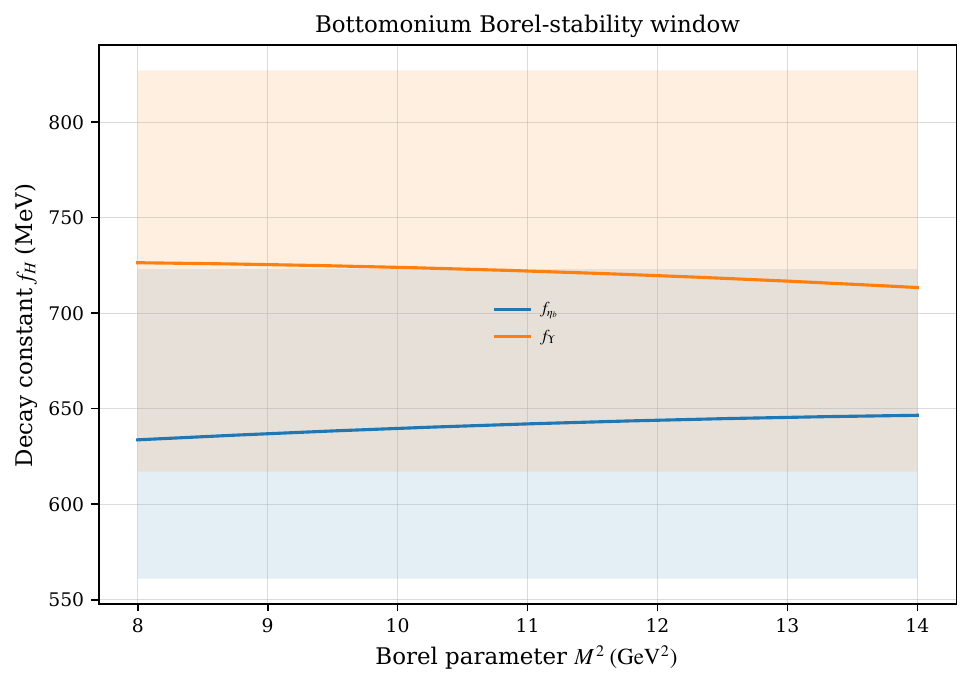}
    \caption{Bottomonium.}
\end{subfigure}
\caption{Representative Borel stability curves for the pseudoscalar and vector residues.  The plotted curves show the central window behavior used to assign the uncertainties in Table~\ref{tab:inputs}.}
\label{fig:borel_plateaus}
\end{figure}

\noindent The QCD sum-rule part of the analysis provides the pole residues and decay constants that normalize the short-distance decay observables.  The extracted values are
\begin{equation}
    f_{\eta_c}=392\pm25~\MeV,
    \qquad
    f_{J/\psi}=403\pm36~\MeV,
    \label{eq:fc_results}
\end{equation}
\begin{equation}
    f_{\eta_b}=642\pm81~\MeV,
    \qquad
    f_{\Upsilon}=722\pm105~\MeV.
    \label{eq:fb_results}
\end{equation}
The hierarchy $f_\Upsilon>f_{\eta_b}>f_{J/\psi}\simeq f_{\eta_c}$ is consistent with the stronger localization of bottomonium and with approximate heavy-quark spin symmetry within each family.  The finite differences between pseudoscalar and vector residues reflect the spin-dependent and radiative effects retained in the sum-rule normalization.  Figure~\ref{fig:borel_plateaus} displays the representative Borel stability curves used to identify the accepted windows and assign the quoted uncertainties.

\noindent The annihilation widths derived from these residues are collected in Table~\ref{tab:annihilation}.  The electromagnetic widths are direct tests of the current normalization, while the gluonic widths carry additional sensitivity to $\alpha_s$ and should be regarded as leading short-distance estimates.  The predicted dileptonic hierarchy is physically transparent: although the $\Upsilon$ decay constant is larger, the bottom-quark charge and the inverse-mass factor suppress the bottomonium dileptonic width relative to $J/\psi$.  Similarly, $\eta_b\to\gamma\gamma$ is much smaller than $\eta_c\to\gamma\gamma$.  This pattern agrees with the general expectations from NRQCD, pNRQCD and Salpeter-based annihilation studies, where short-distance coefficients multiply wave-function or residue factors and QCD radiative corrections can be numerically important \cite{Kwong1988Annihilation,Petrelli1998NLO,Brambilla2003Inclusive,Brambilla2020StronglyCoupled,Kim2005PseudoscalarAnnihilation,Wang2007PwaveAnnihilation,Fu2010Annihilation}.

\begin{table}[H]
\caption{Annihilation widths derived from the two-point QCD sum-rule residues.  Branching fractions are shown only when the total width is used as an input.}
\label{tab:annihilation}
\centering
\resizebox{\textwidth}{!}{%
\begin{tabular}{lccc}
\toprule
State and channel & Width (keV) & Branching fraction & Dominant source of uncertainty \\
\midrule
$\eta_c(1S)\to\gamma\gamma$ & $5.34\pm0.68$ & $(1.68\pm0.22)\times10^{-4}$ & $f_{\eta_c}$ \\
$\eta_c(1S)\to gg$ & $(1.73\pm0.41)\times10^{4}$ & $0.54\pm0.13$ & $f_{\eta_c},\alpha_s$ \\
$\eta_b(1S)\to\gamma\gamma$ & $0.292\pm0.074$ & -- & $f_{\eta_b}$ \\
$\eta_b(1S)\to gg$ & $(1.19\pm0.40)\times10^{4}$ & -- & $f_{\eta_b},\alpha_s$ \\
$J/\psi(1S)\to e^+e^-$ & $5.20\pm0.93$ & $0.0561\pm0.0101$ & $f_{J/\psi}$ \\
$J/\psi(1S)\to\mu^+\mu^-$ & $5.20\pm0.93$ & $0.0561\pm0.0101$ & $f_{J/\psi}$ \\
$J/\psi(1S)\to ggg$ & $60.1\pm21.0$ & $0.65\pm0.23$ & $f_{J/\psi},\alpha_s$ \\
$\Upsilon(1S)\to e^+e^-$ & $1.37\pm0.40$ & $0.0253\pm0.0074$ & $f_{\Upsilon}$ \\
$\Upsilon(1S)\to\mu^+\mu^-$ & $1.37\pm0.40$ & $0.0253\pm0.0074$ & $f_{\Upsilon}$ \\
$\Upsilon(1S)\to\tau^+\tau^-$ & $1.35\pm0.39$ & $0.0251\pm0.0073$ & $f_{\Upsilon}$ \\
$\Upsilon(1S)\to ggg$ & $46.0\pm20.4$ & $0.85\pm0.38$ & $f_{\Upsilon},\alpha_s$ \\
\bottomrule
\end{tabular}}
\end{table}

\subsection{Transition form factors and medium-related QCD sum-rule extensions}
\begin{figure}[H]
\centering
\begin{subfigure}{0.48\textwidth}
    \centering
    \includegraphics[width=\linewidth]{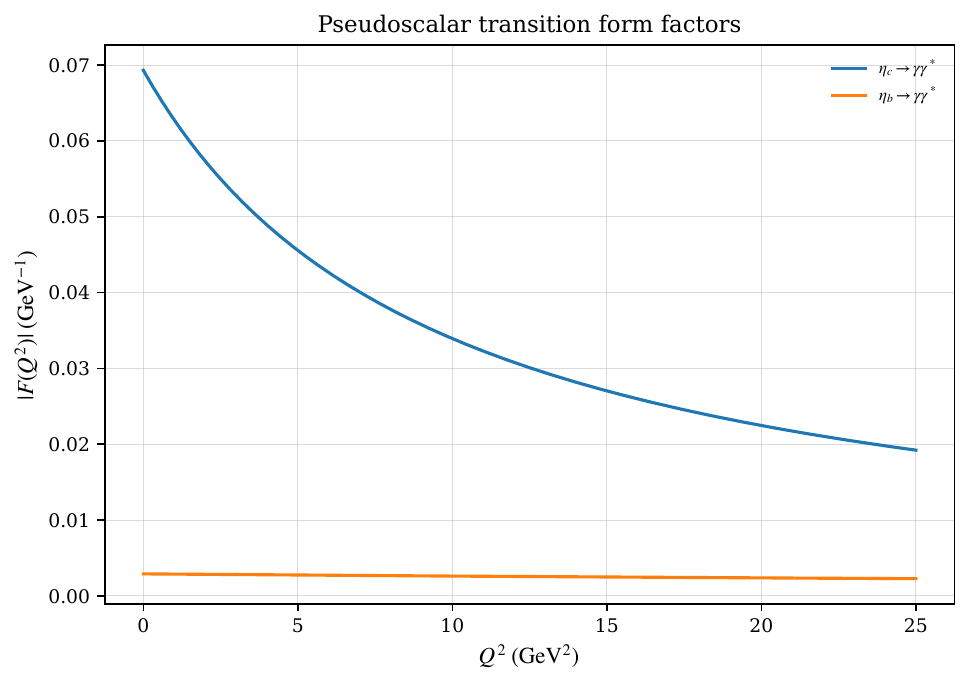}
    \caption{$P\to\gamma\gamma^*$.}
\end{subfigure}\hfill
\begin{subfigure}{0.48\textwidth}
    \centering
    \includegraphics[width=\linewidth]{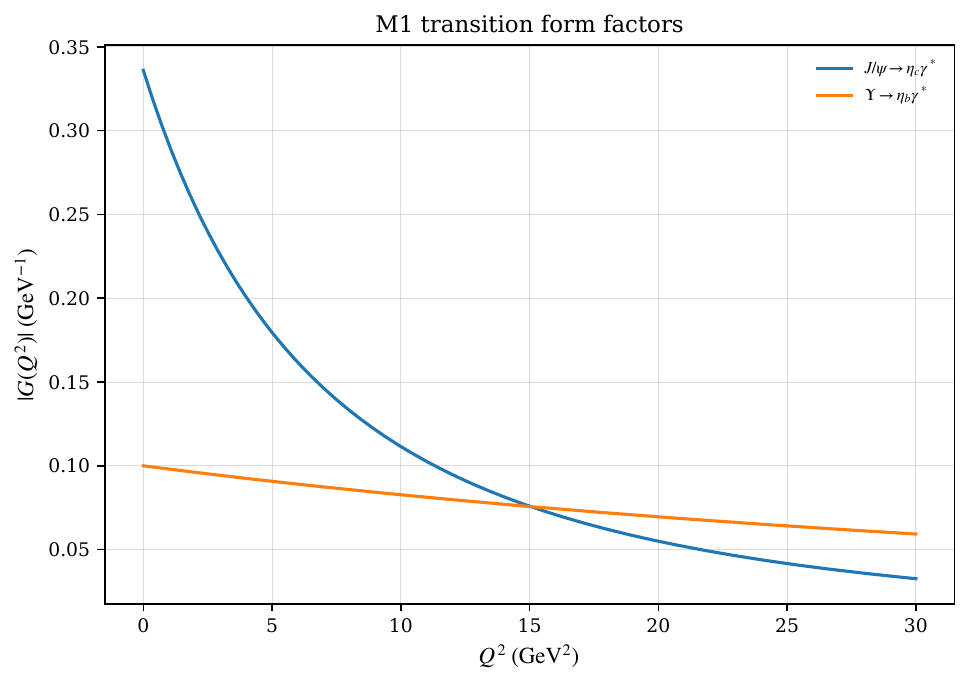}
    \caption{$V\to P\gamma^*$.}
\end{subfigure}
\caption{Spacelike transition form factors used for radiative decays.  The plotted curves show the absolute form factors in GeV$^{-1}$; the accompanying CSV file also gives the normalized ratios.}
\label{fig:form_factors}
\end{figure}

\noindent The transition-form-factor parameters in Table~\ref{tab:ff} give a compact representation of the spacelike QCD sum-rule output used for the electromagnetic curves plotted in Fig.~\ref{fig:form_factors}.  The pseudoscalar two-photon form factors are represented by a monopole form, while the $V\to P\gamma^*$ M1 form factors are represented by a dipole form.  The normalizations are larger in charmonium because the relevant mass scale is lower and the charm-quark charge is larger.  Bottomonium form factors decrease more slowly with $Q^2$ because the pole scales are larger, but their values at $Q^2=0$ are smaller.

\begin{table}[H]
\caption{Transition form-factor parameters used for the plotted curves.  A monopole fit is used for $P\to\gamma\gamma^*$ and a dipole fit for $V\to P\gamma^*$.}
\label{tab:ff}
\centering
\resizebox{\textwidth}{!}{%
\begin{tabular}{lcccc}
\toprule
Transition & Fit & $F(0)$ or $G(0)$ (GeV$^{-1}$) & $\Lambda$ (GeV) & Radius (fm) \\
\midrule
$\eta_c(1S)\to\gamma\gamma^*$ & monopole & $0.0693\pm0.0044$ & 3.0969 & 0.156 \\
$\eta_b(1S)\to\gamma\gamma^*$ & monopole & $0.00290\pm0.00037$ & 9.4603 & 0.051 \\
$J/\psi(1S)\to\eta_c(1S)\gamma^*$ & dipole & $0.336\pm0.038$ & 3.6861 & 0.185 \\
$\Upsilon(1S)\to\eta_b(1S)\gamma^*$ & dipole & $0.100\pm0.030$ & 10.0233 & 0.068 \\
\bottomrule
\end{tabular}}
\end{table}

\begin{figure}[H]
\centering
\includegraphics[width=0.62\textwidth]{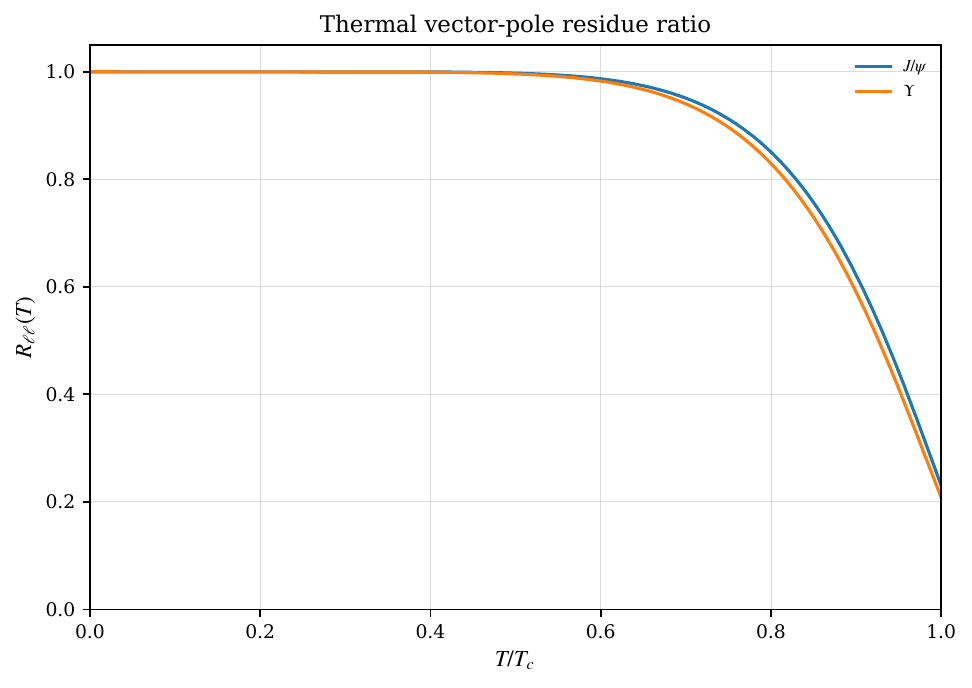}
\caption{Thermal suppression of the vector dileptonic pole-strength ratio.  The parametrization is normalized to reproduce the limiting behavior reported by the finite-temperature vector-current QCD sum rule.}
\label{fig:thermal_residue}
\end{figure}

\noindent The thermal vector residue, finite-momentum energy-shift envelope and spin decomposition are included as controlled QCD sum-rule extensions rather than as new independent spectroscopy inputs.  Here $T_c$ denotes the pseudo-critical temperature scale used in the finite-temperature QCD sum-rule inputs.  The thermal residue plot in Fig.~\ref{fig:thermal_residue} uses the finite-temperature vector-current result to show that the dileptonic pole strength is suppressed mainly through $f_V(T)$ near $T_c$.  The illustrative ratios are $R_{\ell\ell}(T_c)=0.230$ for $J/\psi$ when $M_V(T_c)/M_V(0)=0.88$, and $R_{\ell\ell}(T_c)=0.208$ for $\Upsilon$ when $M_V(T_c)/M_V(0)=0.975$ \cite{Veliev2011Vector}.

\noindent The finite-momentum estimates in Table~\ref{tab:momentum} show a much stronger potential effect for charmonium than for bottomonium.  A $-3\%$ charmonium mass shift at $|\mathbf q|=1~\GeV$ corresponds to an energy shift of order $85$--$90~\MeV$, whereas the bottomonium shift remains below $1~\MeV$ in the quoted envelope.  This difference reflects the greater robustness of the compact bottomonium ground states and the stronger medium sensitivity of charmonium pole positions near $T_c$ \cite{Kim2023Momentum}.

\begin{table}[H]
\caption{Finite-momentum in-medium energy-shift envelope near $T_c$.}
\label{tab:momentum}
\centering
\begin{tabular}{lcccc}
\toprule
State & $|\mathbf q|$ (GeV) & $\delta M/M$ (\%) & $E_{\rm vac}$ (GeV) & $\Delta E$ (MeV) \\
\midrule
$\eta_c(1S)$ & 1.0 & $-3.0$ & 3.147 & $-84.9$ \\
$J/\psi(1S)$ & 1.0 & $-3.0$ & 3.254 & $-88.4$ \\
$\eta_b(1S)$ & 4.0 & $-0.01$ & 10.215 & $-0.86$ \\
$\Upsilon(1S)$ & 4.0 & $-0.01$ & 10.276 & $-0.87$ \\
\bottomrule
\end{tabular}
\end{table}

\noindent The rotating-frame decomposition in Table~\ref{tab:spin} further clarifies the spin-one current content.  The vector states are dominated by the quark-spin contribution, especially in bottomonium, while the axial-vector states carry a much larger orbital component.  This observation is useful because the same vector and axial-vector currents enter the decay constants and pole residues discussed above.  The finite-temperature spin-structure analysis indicates that the quark-spin and orbital pieces change moderately near $T_c$, while the gluonic contribution remains small \cite{Kim2025SpinThermal}.

\begin{table}[H]
\caption{Spin decomposition of spin-one quarkonia from the rotating-frame QCD sum-rule analysis.  Small deviations from exactly 100\% arise from rounding the tabulated components.}
\label{tab:spin}
\centering
\begin{tabular}{lcccc}
\toprule
State & $S_q$ (\%) & $L_q$ (\%) & $L_p$ (\%) & $J_g$ (\%) \\
\midrule
$J/\psi$ & 88.1 & 11.4 & 0.20 & 0.80 \\
$\Upsilon(1S)$ & 92.7 & 7.6 & 0.004 & 0.015 \\
$\chi_{c1}$ & 40.8 & 61.5 & 0.082 & $-1.5$ \\
$\chi_{b1}$ & 43.1 & 57.0 & $-0.001$ & $-0.005$ \\
\bottomrule
\end{tabular}
\end{table}

\section{Conclusions}\label{sec:conclusions}
\noindent This work has presented a unified hidden-flavour study of charmonium and bottomonium in which the spectrum, radiative transitions, QCD sum-rule observables, finite-spectrum thermodynamic indicators and Regge systematics are treated as connected parts of the same phenomenological framework.  The calculation uses a screened Coulomb-plus-confining interaction with spin-dependent corrections for the mass spectrum, while two-point and three-point QCD sum rules are used to organize the decay constants, annihilation widths and electromagnetic transition form factors.  The aim is not to make separate model comparisons for each observable, but to test whether the same $c\bar c$ and $b\bar b$ state assignments lead to a consistent pattern across spectroscopy, decays and global trajectory analyses.

\noindent The combined numerical results support a clear physical picture.  The low-lying mass spectra fix the short-distance and intermediate-distance parts of the interaction and remain close to the established charmonium and bottomonium levels, whereas the higher radial and orbital states are more sensitive to screening, spin splittings and possible threshold effects.  The Regge trajectories provide a global check on this ordering.  With $M^2$ on the horizontal axis and either $J^{PC}$ or $n_r$ on the vertical axis, the calculated states show the expected approximate linearity.  The $b\bar b$ trajectories are more compressed than the $c\bar c$ ones, reflecting the larger bottom-quark mass and the smaller level spacings of bottomonium.

\noindent The E1 and M1 transition widths give a more differential test of the wave functions.  The E1 transitions are governed by photon phase space, angular coefficients and radial overlap integrals, and therefore probe the consistency of the orbital assignments across the $S$-, $P$-, $D$- and higher-wave sectors.  The M1 transitions test spin-flip dynamics and radial-node cancellations; their stronger suppression in bottomonium follows naturally from the heavier quark mass and the compactness of the $b\bar b$ wave functions.  The contrast between allowed and hindered channels is therefore an important internal diagnostic of the spin and radial structure of the calculated states.

\noindent The QCD sum-rule part of the study complements the potential-based spectrum by connecting the same hidden-flavour states to current-coupled observables.  The pseudoscalar and vector decay constants, annihilation widths and transition form factors are controlled by pole residues, continuum thresholds and Borel stability.  Thermal-residue, finite-momentum and rotating-frame extensions are retained only as controlled QCD sum-rule diagnostics, not as substitutes for a genuine in-medium quarkonium calculation.  The finite-spectrum thermodynamic observables are treated in the same conservative way: by using excitation energies relative to the calculated ground state, the free energy, mean excitation energy, entropy and specific heat summarize the level density and spin degeneracy of the vacuum spectrum rather than the bulk thermodynamics of a hot QCD medium.

\noindent Taken together, the repeated hierarchy across independent observables is the main outcome of the work.  Bottomonium is consistently more compact, more compressed in Regge slope and more suppressed in spin-flip radiative transitions, while charmonium shows stronger sensitivity to fine-structure splittings, radial-node effects and medium-related current-correlator modifications.  The calculation is therefore internally consistent because the same physical pattern reappears in the mass tables, Regge plots, E1/M1 widths, decay constants, annihilation widths, transition form factors and finite-spectrum thermal indicators.

\noindent The main limitations are also clear.  Coupled-channel effects near open-flavour thresholds, strong state mixing in the higher charmonium region, continuum-induced mass shifts and genuine finite-temperature or in-medium modifications are not included explicitly.  These effects are expected to be most important for states close to open-charm thresholds and for highly excited levels.  Future extensions should therefore combine the present vacuum framework with coupled-channel dynamics, improved lattice and experimental inputs, and a more complete finite-temperature treatment.  Within its stated scope, however, the present study provides a coherent hidden-charm and hidden-bottom benchmark for spectroscopy, radiative transitions and QCD sum-rule observables.


\end{document}